\documentclass[12pt,preprint]{aastex}
\def\degpoint{\ifmmode ^{\rm{o}}\!. \else $^{\rm{o}}\!.$\fi}

\newcommand{\ms}{\mbox{m\,s$^{-1}$}}

\newcommand{\Msun}{\mbox{M$_{\odot}$}}

\newcommand{\Mjup}{\mbox{M$_{\rm Jup}$}}

\newcommand{\Mearth}{\mbox{M$_{\oplus}$}}

\newcommand{\ltsimeq}{\raisebox{-0.6ex}{$\,\stackrel
         {\raisebox{-.2ex}{$\textstyle <$}}{\sim}\,$}}
\newcommand{\gtsimeq}{\raisebox{-0.6ex}{$\,\stackrel
         {\raisebox{-.2ex}{$\textstyle >$}}{\sim}\,$}}

\begin{document}

\title{The Frequency of Low-Mass Exoplanets. II. The `Period Valley' }

\author{Robert A.~Wittenmyer\altaffilmark{1}, Simon 
J.~O'Toole\altaffilmark{2}, H.R.A.~Jones\altaffilmark{3}, 
C.G.~Tinney\altaffilmark{1}, R.P.~Butler\altaffilmark{4}, 
B.D.~Carter\altaffilmark{5}, J.~Bailey\altaffilmark{1} }
\altaffiltext{1}{Department of Astrophysics, School of Physics, 
University of NSW, 2052, Australia}
\altaffiltext{2}{Anglo-Australian Observatory, PO Box 296, Epping, 1710, 
Australia}
\altaffiltext{3}{Centre for Astrophysics Research, University of 
Hertfordshire, College Lane, Hatfield, Herts AL10 9AB, UK}
\altaffiltext{4}{Department of Terrestrial Magnetism, Carnegie 
Institution of Washington, 5241 Broad Branch Road, NW, Washington, DC 
20015-1305, USA}
\altaffiltext{5}{Faculty of Sciences, University of Southern Queensland, 
Toowoomba, Queensland 4350, Australia}
\email{
rob@phys.unsw.edu.au}

\shorttitle{The Frequency of Low-Mass Exoplanets. II.}
\shortauthors{Wittenmyer et al.}

%  Hugh's comments added 18 Jan
%-------------------------------------------------------------------
\begin{abstract}

\noindent Radial-velocity planet search campaigns are now beginning to 
detect low-mass ``Super-Earth'' planets, with minimum masses M~sin~$i 
\ltsimeq$ 10\Mearth.  Using two independently-developed methods, we have 
derived detection limits from nearly four years of the highest-precision 
data on 24 bright, stable stars from the Anglo-Australian Planet Search.  
Both methods are more conservative than a human analysing an individual 
observed data set, as is demonstrated by the fact that both techniques 
would detect the radial velocity signals announced as exoplanets for the 
61~Vir system in 50\% of trials.  There are modest differences between 
the methods which can be recognised as arising from particular criteria 
that they adopt.  What both processes deliver is a \textit{quantitative} 
selection process such that one can use them to draw quantitative 
conclusions about planetary frequency and orbital parameter distribution 
from a given data set.  Averaging over all 24 stars, in the period range 
$P<$300 days and the eccentricity range $0.0<e<0.6$, we could have 
detected 99\% of planets with velocity amplitudes $K\gtsimeq$7.1 \ms.  
For the best stars in the sample, we are able to detect or exclude 
planets with $K\gtsimeq$3 \ms, corresponding to minimum masses of 
8\Mearth\ (P=5 days) or 17\Mearth\ (P=50 days).  Our results indicate 
that the observed ``period valley,'' a lack of giant planets ($M>100$ 
M$_{\oplus}$) with periods between 10-100 days, is indeed real.  However, 
for planets in the mass range 10-100 M$_{\oplus}$, our results suggest 
that the deficit of such planets may be a result of selection effects.

\end{abstract}

\keywords{planetary systems -- techniques: radial velocities }

%--------------------------------------------------------------------
\section{Introduction}

Extrasolar planet detections are inexorably pushing toward ever-lower 
masses, and it seems that each new mass threshold spawns an unavoidably 
anthropocentric moniker.  First, in 1995, were the ``hot Jupiters'' (or 
``roasters''; Sudarsky et al.~2000), with minimum masses m~sin~$i$ 
greater than the mass of Jupiter \citep{mayor95}.  The earliest planet 
discoveries fell into this category due to the well-known 
radial-velocity selection bias toward detecting massive planets in 
short-period orbits ($P\ltsimeq$10 days).  The discovery of several 
Neptune-mass (=17.4 \Mearth) planets in 2004 \citep{mcarthur04, 
santos04, bonfils05} brought the ``hot Neptune'' into common parlance.  
Most recently, the discovery of planets with m~sin~$i$ less than 10 
Earth masses has prompted the somewhat misleading appellation of 
``super-Earth.'' These planets, of course, are not likely to bear any 
resemblance to Earth, and models of the composition of such objects 
strongly suggest that almost all of these worlds possess thick 
atmospheres surrounding a core of rock and ice \citep{seager07,adams08}.

%  we say super-earth to avoid saying mini-uranus.

The Anglo-Australian Planet Search (AAPS) has been in operation at the 
3.9m Anglo-Australian Telescope for 11 years, and has published 36 
exoplanet discoveries.  By integrating for at least 20 minutes on each 
target in order to average over the stellar p-mode oscillations 
\citep{otoole08}, the AAPS now achieves velocity precision approaching 1 
\ms\ for bright, inactive stars \citep{monster}.  Since 2005~June, the 
AAPS has been routinely implementing this observing strategy, which has 
(together with long continuous observing blocks) proven instrumental in 
our recent discovery of a super-Earth and two Neptune-mass planets 
orbiting 61~Vir \citep{61vir}.

In \citet{monster}, we explored the frequency of planets with periods 
less than $\sim$20 days.  In that work, we derived detection 
sensitivities from a continuous 48-night observing campaign targeting 24 
bright, stable stars.  The stars have spectral types between G0 and K5, 
and they range in mass from 0.77-1.28 \Msun; complete details are given 
in Table~1 of \citet{monster}.  In this paper, we consider periods 
ranging up to 300 days, now making use of nearly 4 years of high-quality 
data on these 24 stars from the AAPS.  The observational data are 
summarised in Table~1, and the parameters of the five planets which have 
been found in this sample are given in Table~\ref{planets}.  We also 
bring two independent simulation algorithms to bear on these data to 
derive extremely tight limits on planetary companions in the range 
$2d<P<300d$.  The planetary periods and masses probed by these data are 
interesting for two reasons.  First, this region includes the ``period 
valley'' noted by \citet{jones03} and \citet{udry03} -- a lack of 
planets with orbital periods of 10-100 days.  Second, the large amount 
of high-precision data used here enables the detection of potentially 
terrestrial-mass planets.  Core-accretion simulations which include 
Type~I migration \citep{idalin04a, idalin08b} predict an abundance of 
planets with masses less than about 10 \Mearth\ orbiting within 1~AU, 
and a lack of planets with higher masses in that region (the ``planet 
desert'').  The large, extremely high-precision data set analysed here 
represents one of the first able to test these predictions.

In Section 2, we describe the two methods used to derive detection 
limits from these data.  Section~3 presents the results and discusses 
differences between the two methods.  In Section~4, we further discuss 
the implications of our results with regards to the underlying 
distribution of exoplanets.

%--------------------------------------------------------------------
\section{Computational Methods}

There are two ways to approach the input data for a detection-limit 
computation: using the actual radial-velocity data \citep{limitspaper, 
endl02}, or creating simulated data sets using the observation dates and 
uncertainties of the real data \citep{otoole09a, cumming99}.  We have 
employed independent methods using each of these approaches to determine 
the types of planets to which these data were sensitive.

\subsection{Simulated Data}

The first simulation technique, described fully in \citet{otoole09a}, 
seeks to recover simulated Keplerian signals (with appropriate noise 
terms arising from measurement uncertainties and stellar jitter) sampled 
at the epochs of the actual data and retaining the original 
uncertainties of the actual data.  The simulated velocity curves are 
analysed using the two-dimensional Lomb-Scargle periodogram 
\citep{otoole07,otoole09a} which extends the standard Lomb-Scargle 
periodogram to search eccentricity as well as period space.

The simulated observations for this study were constructed in the 
following way. We used a grid of periods, eccentricities and planet 
masses as input in the same way as \citet{otoole09a}, with periods in 
the range 2--320 days, eccentricities from 0.0 to 0.6 in steps of 0.1 
and masses of 0.005, 0.01, 0.02, 0.05, 0.1, 0.2, 0.5 and 1.0 M$_J$. Each 
set of parameters was simulated 100 times, leading to 72800 simulated 
velocity curves for each of the 24 stars.

This approach does not use the standard $F$-test to determine the 
reliability of a detection. Instead, it uses four detection criteria as 
follows \citep{otoole09a}.  First, the RMS of the simulated observations 
must be greater than or equal to the RMS of the residuals of the 
best-fitting model to those data.  That is, the scatter should decrease 
after the subtraction of a Keplerian model.  Second, the measured 
semi-amplitude must be greater than or equal to twice the measured 
semi-amplitude uncertainty plus the RMS of the residuals of the 
best-fitting model. Third, the measured period must be greater than or 
equal to twice the measured period uncertainty. Finally, the value of 
$\chi^2_\nu$ must be less than or equal to three. The last criterion is 
only useful when stellar jitter is not included, as discussed by 
\citet{otoole09a}; in this paper and \citet{monster} -- where jitter is 
included -- it is somewhat redundant. The other three criteria are a 
gauge of the reliability of a fit.

With these criteria, the method does not rely on the assumptions of the 
$F$-test. Its chief strengths lie in its ability to recover 
significantly eccentric orbits \citep{otoole09a} which would be 
undetectable using a standard Lomb-Scargle periodogram.  In addition, 
because this method performs a full Keplerian fit to the simulated data, 
the distribution of orbital parameters derived from the fits provides 
valuable information on the biases inherent in fitting radial-velocity 
data \citep{otoole09a}. For the rest of this paper we will refer to this 
algorithm as Method~1.

\subsection{Real Data}

The second technique, described fully in \citet{limitspaper}, adds the 
Keplerian signal of a simulated planet to the original data, and then 
attempts to recover that signal using a Lomb-Scargle periodogram 
\citep{lomb76, scargle82}.  Previous work using similar methodology 
includes \citet{walker95} and \citet{endl02}.  We will refer to this 
algorithm as Method~2.  In this work, we required that the recovered 
period be within 5\% of the injected period, and that its false-alarm 
probability (FAP) be less than 0.1\%.  As the null hypothesis of this 
method is that the data are purely noise and contain no signals due to 
real planets, the known planets HD~4308b \citep{udry06} and HD~16417b 
\citep{16417paper} must be treated differently.  As noted in 
\citet{wittenmyer09}, the velocity signature of an unknown planet can be 
altered by the presence of an additional (known) planet.  Hence, we have 
determined the detection limits for HD~4308 and HD~16417 using the 
method of \citet{wittenmyer09} wherein the simulated planet is added 
prior to fitting for and removing the known planet's signal.  For the 
Alpha Centauri system (HD~128620/1), we removed a quadratic fit to the 
data, which provided a satisfactory approximation of the binary orbit 
over the short time period considered here.  We also removed linear 
trends from the long-period binaries HD~10360 and HD~10361.  Each star 
received 7 trials at the 99\% recovery level, for trial signals with 
$e=$0.0, 0.1, 0.2, 0.3, 0.4, 0.5, and 0.6.  In addition, simulations at 
90\%, 70\%, 50\%, and 10\% recovery were performed at $e=0.0$ for each 
star.

This method is computationally faster than Method~1, since each trial 
only requires the computation of a simple periodogram, and there are 
considerably fewer such trials.  Method~2 also has the advantage that 
the noise characteristics of the real data are preserved, whereas for 
simulated data, additional assumptions must be made about the noise 
term.  However, it breaks down when there are fewer than $N\sim$30 data 
points, as the reliability of the FAP calculation is strongly dependent 
on the number of data points.  With limited data, even a very strong 
periodogram peak may fail to reach the 0.1\% FAP level.  This is less 
relevant for the current work, as all 24 stars considered here have at 
least 31 observations (Table~\ref{rvdata1}).  We note that 
\citet{cumming08} in their Appendix give formulae for the FAP 
calculation in the regime of small $N$.  For future work, the FAP 
calculations in Method~1 can be modified in this manner to better handle 
small $N$.

%--------------------------------------------------------------------
\section{Results}

\subsection{Detectabilities}

\begin{figure}
\plotone{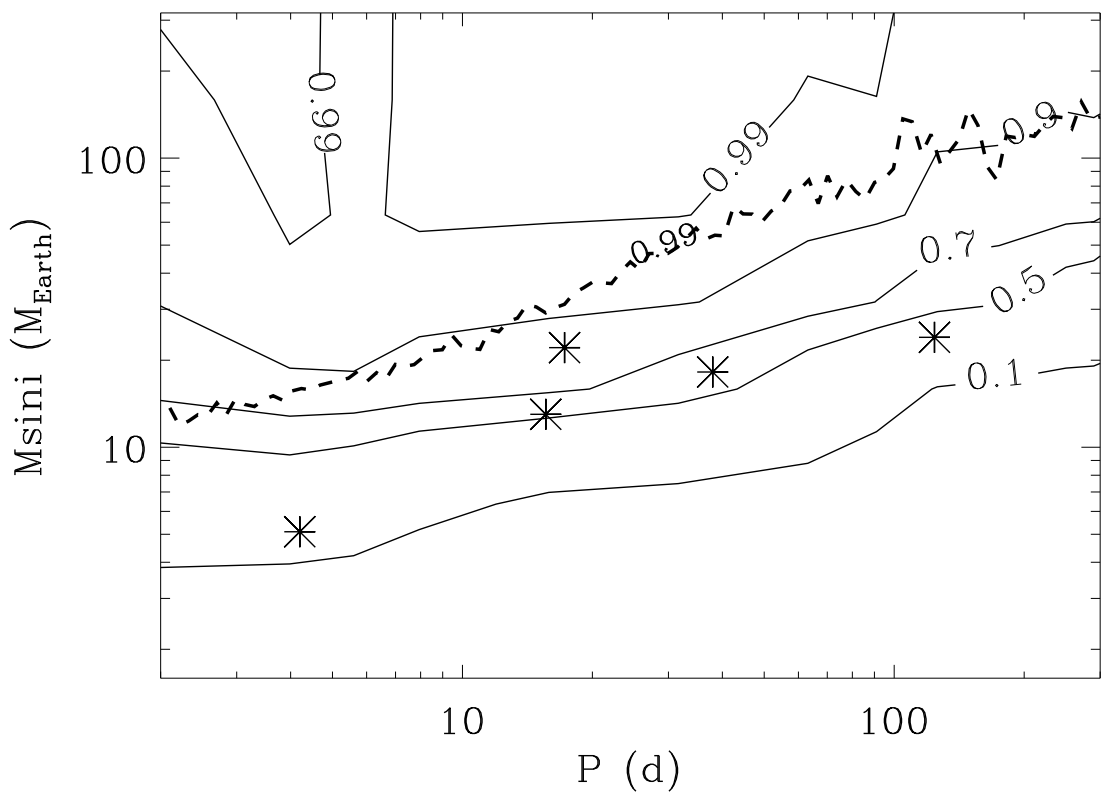}
\caption{Detectabilities obtained from Method~1 (solid contours) 
averaged over all 24 stars, for planets with eccentricities from 0.0 to 
0.6.  Contours indicate the fraction of injected planets that were 
recovered.  Results from Method~2 (99\% recovery) for the same range of 
eccentricities are shown as a dotted line.  The five announced planets 
in this sample are plotted as large asterisks. }
\label{means}
\end{figure}

\begin{figure}
\plottwo{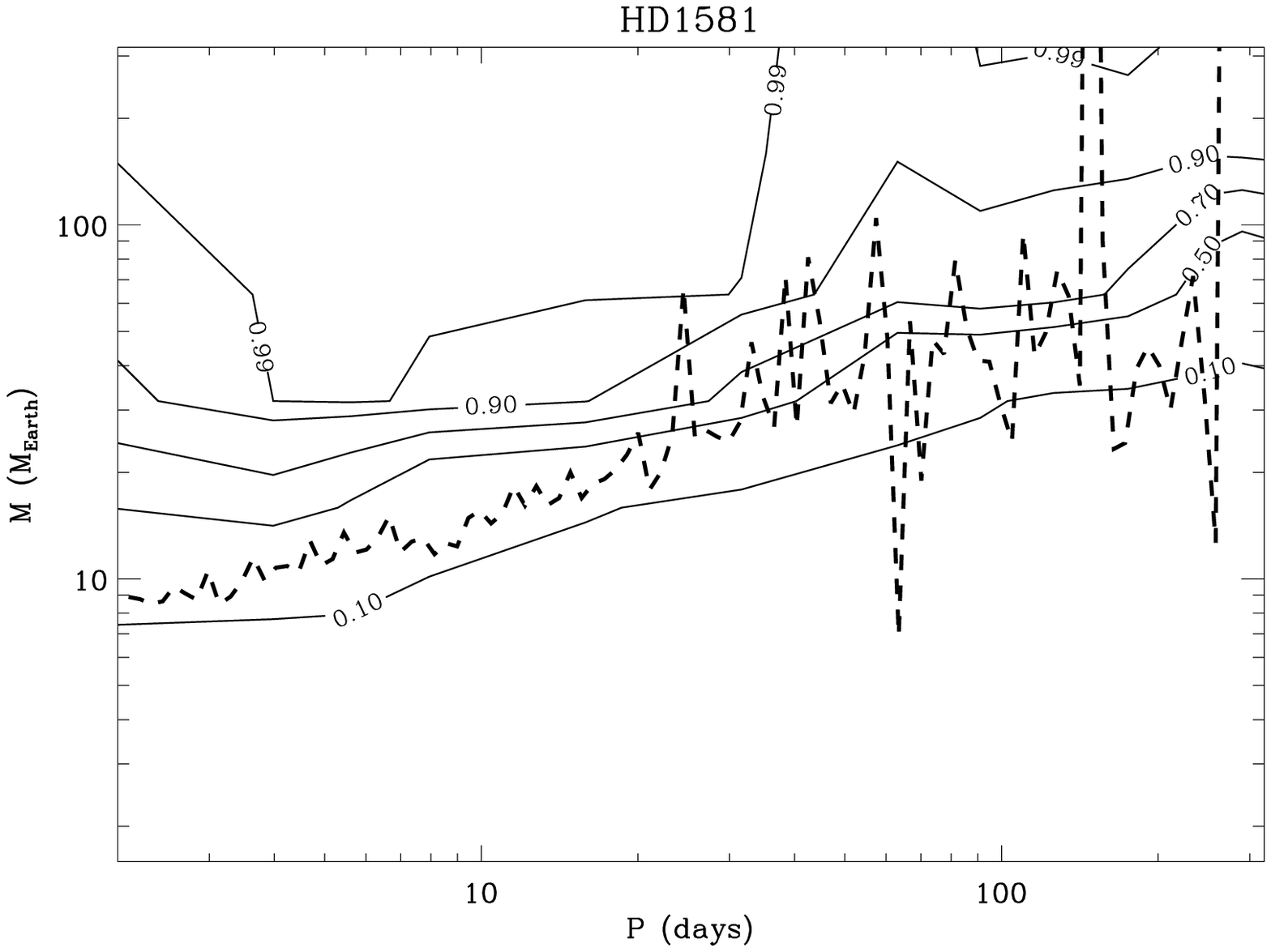}{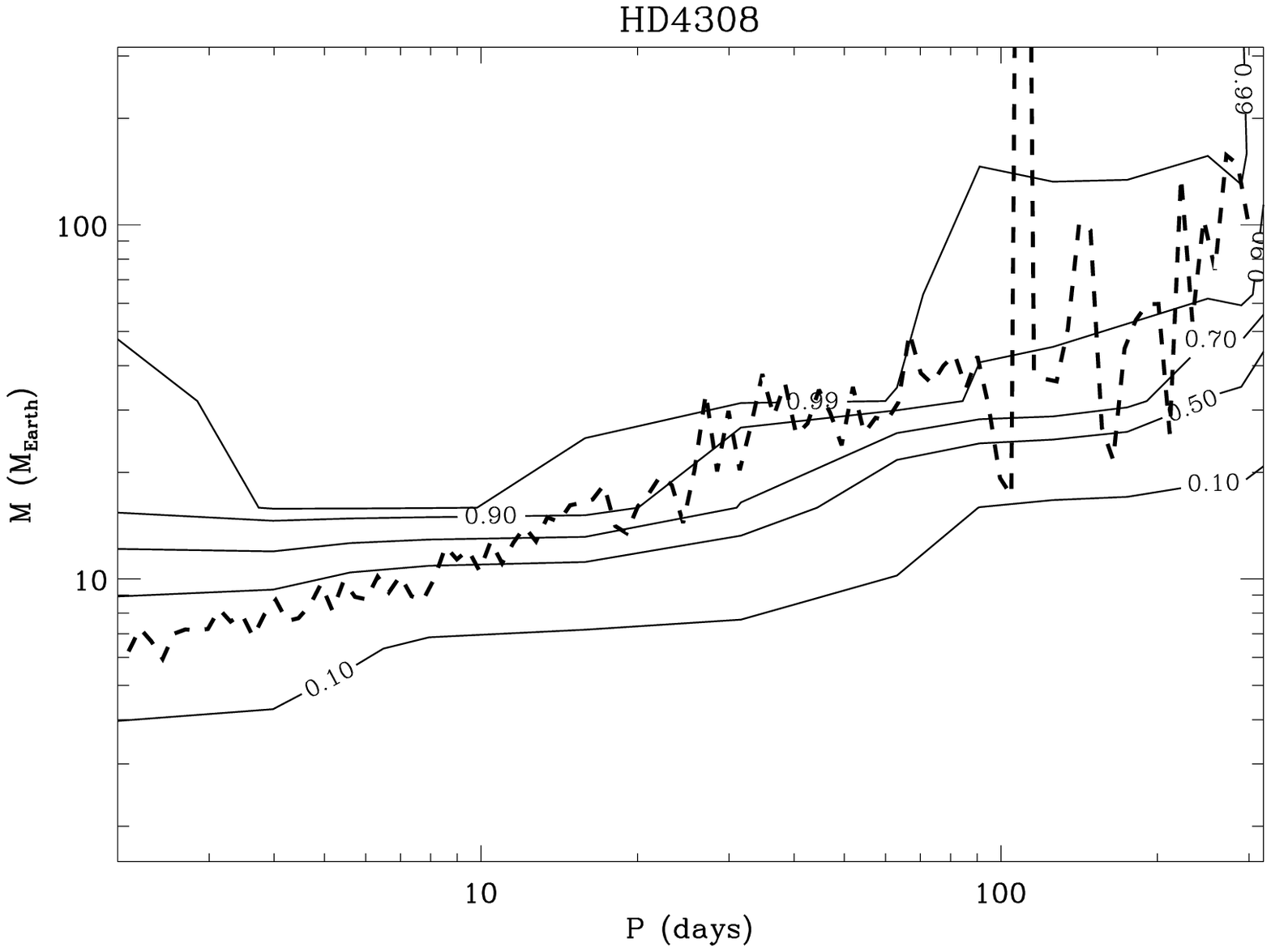}
\caption{Detectabilities obtained from Method~1 (solid contours), for 
HD~1581 (left panel) and HD~4308 (right panel), for planets with 
eccentricities from 0.0 to 0.6.  Contours indicate the fraction of 
injected planets that were recovered.  Results from Method~2 (99\% 
recovery) for the same range of eccentricities are overplotted as a 
dotted line. }
\label{results1}
\end{figure}

\begin{figure}
\plottwo{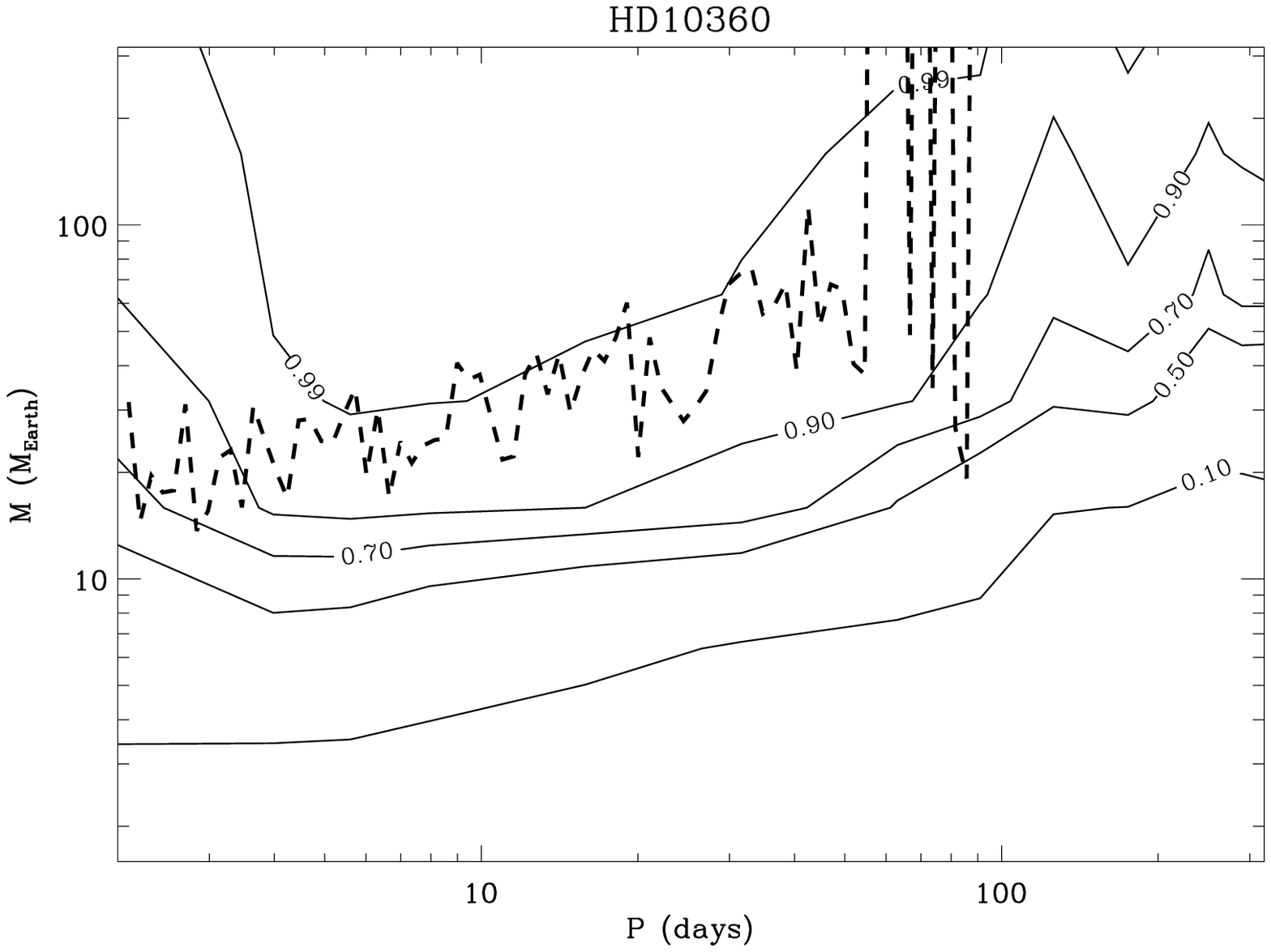}{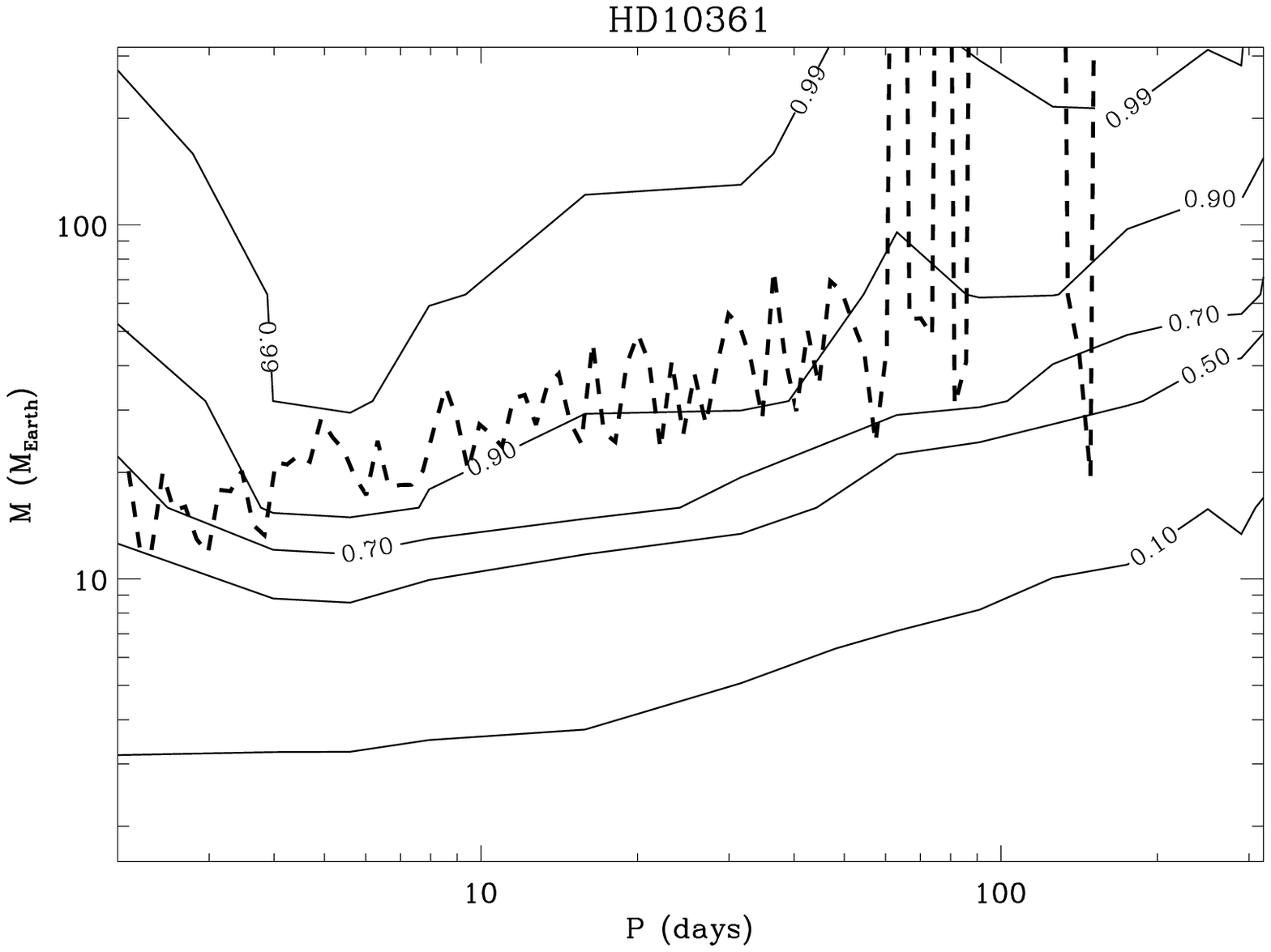}
\caption{Same as Figure~\ref{results1}, but for HD~10360 (left) and 
HD~10361 (right). }
\label{results2}
\end{figure}

\begin{figure}
\plottwo{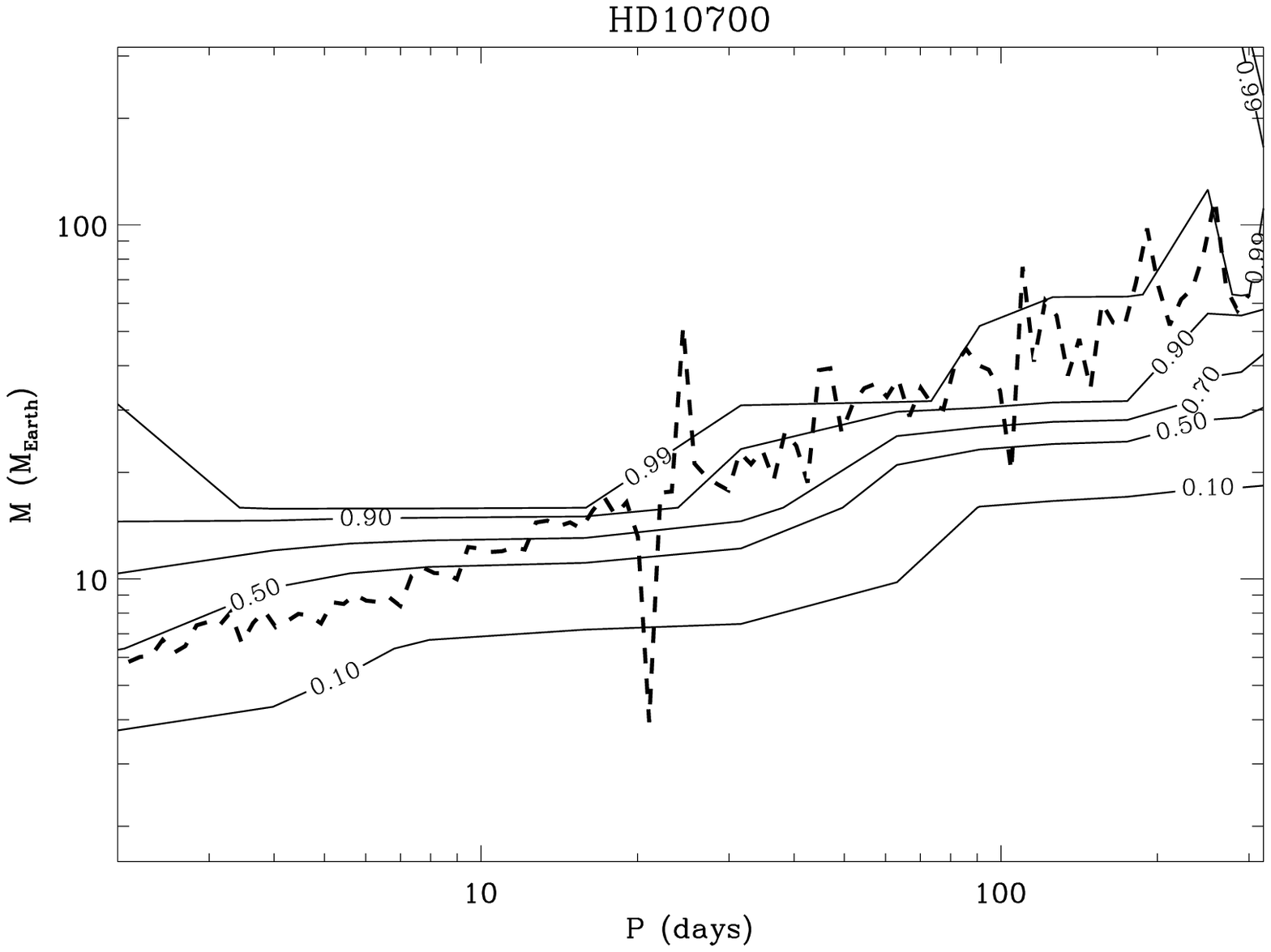}{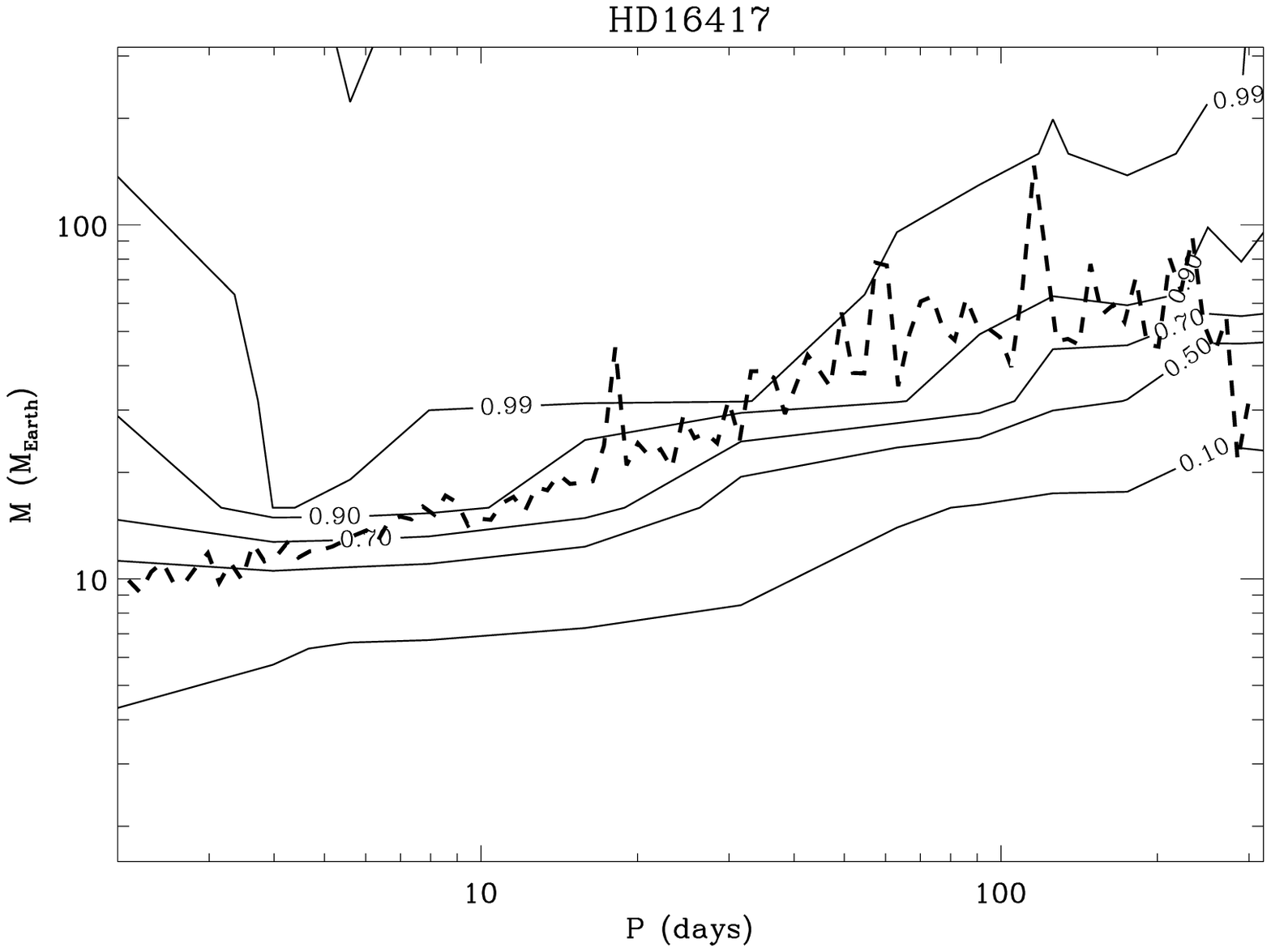}
\caption{Same as Figure~\ref{results1}, but for HD~10700 (left) and 
HD~16417 (right). }
\label{results3}
\end{figure}

\begin{figure}
\plottwo{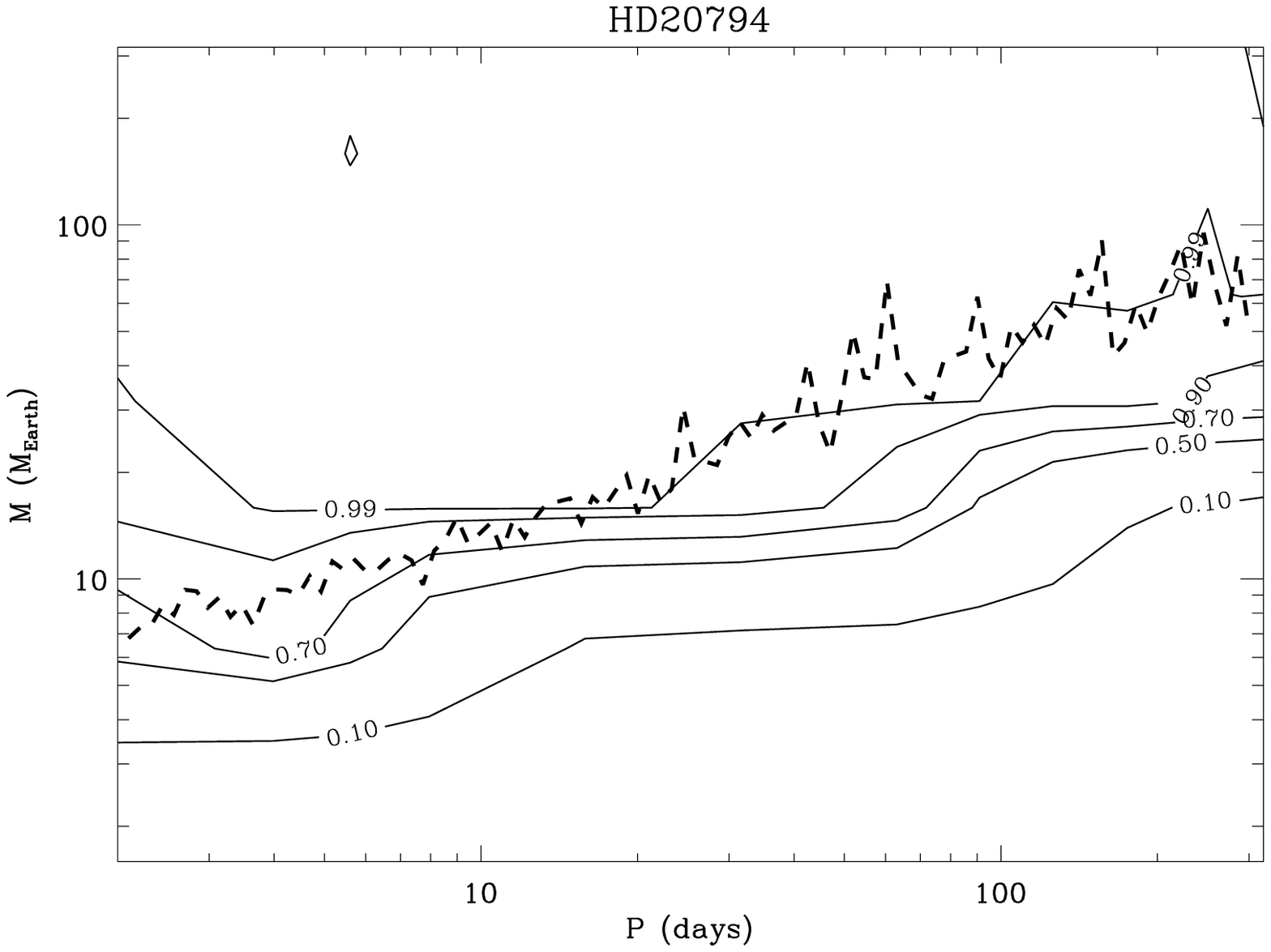}{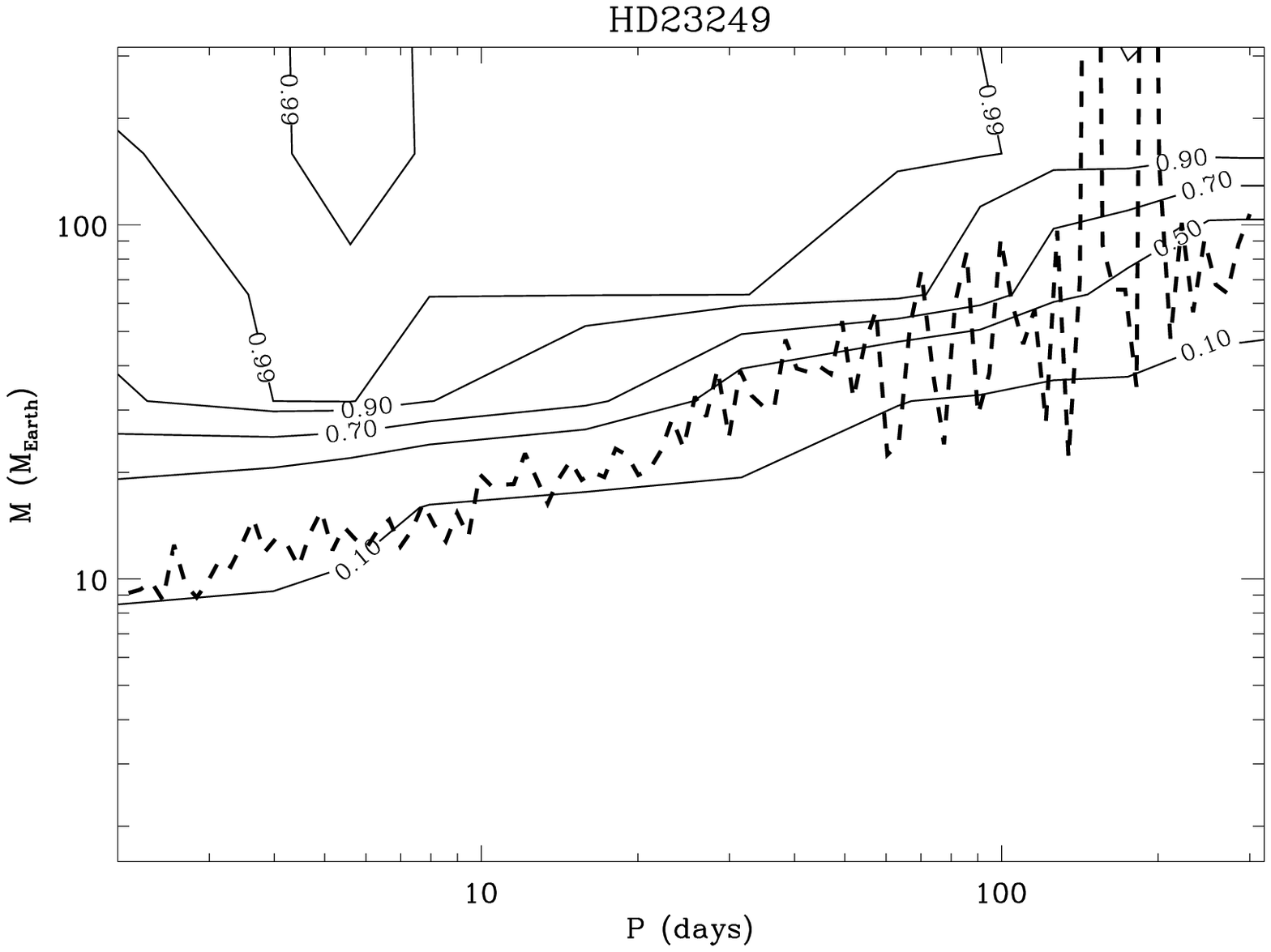}
\caption{Same as Figure~\ref{results1}, but for HD~20794 (left) and 
HD~23249 (right). }
\label{results4}
\end{figure}

\begin{figure}
\plottwo{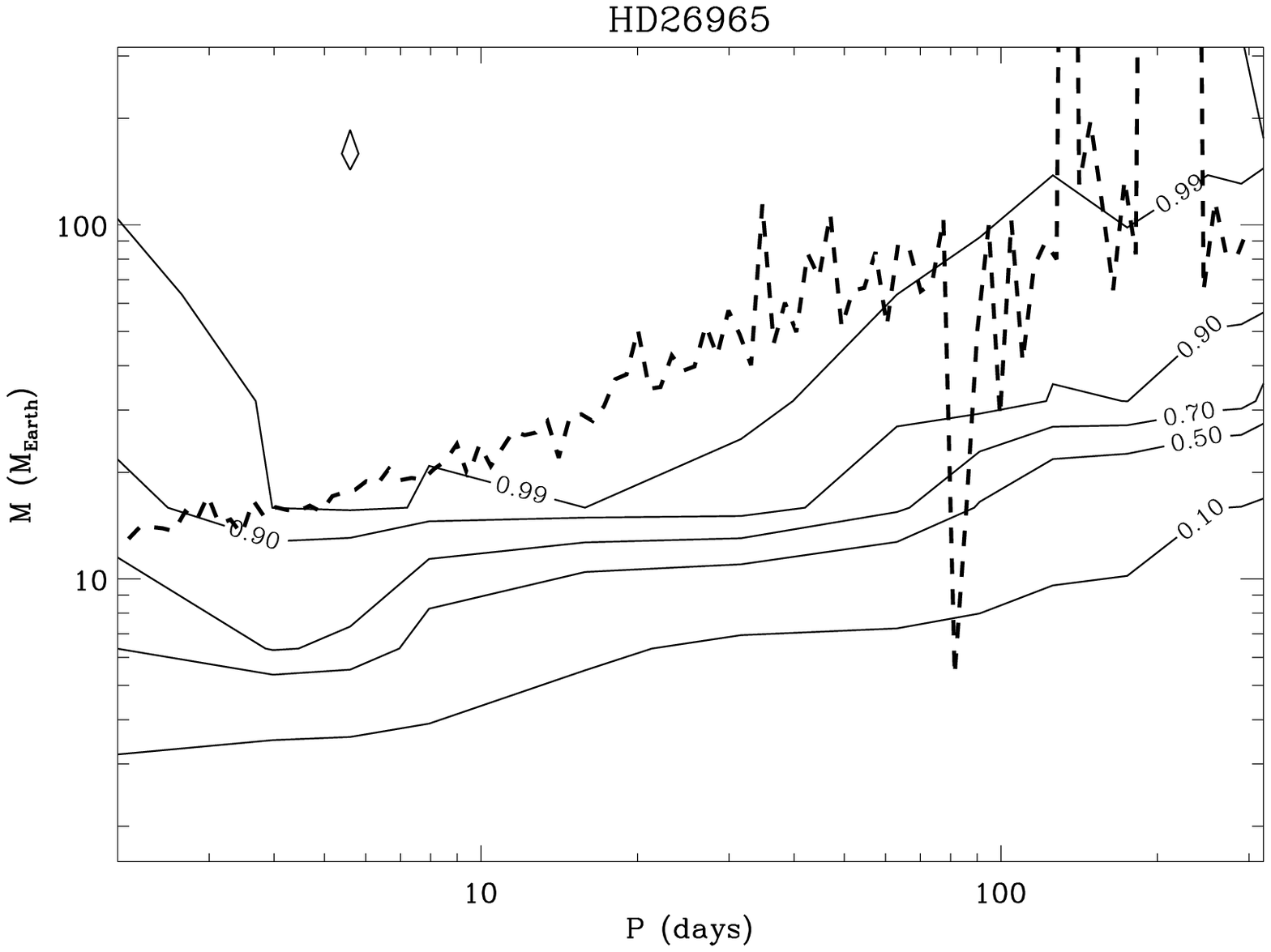}{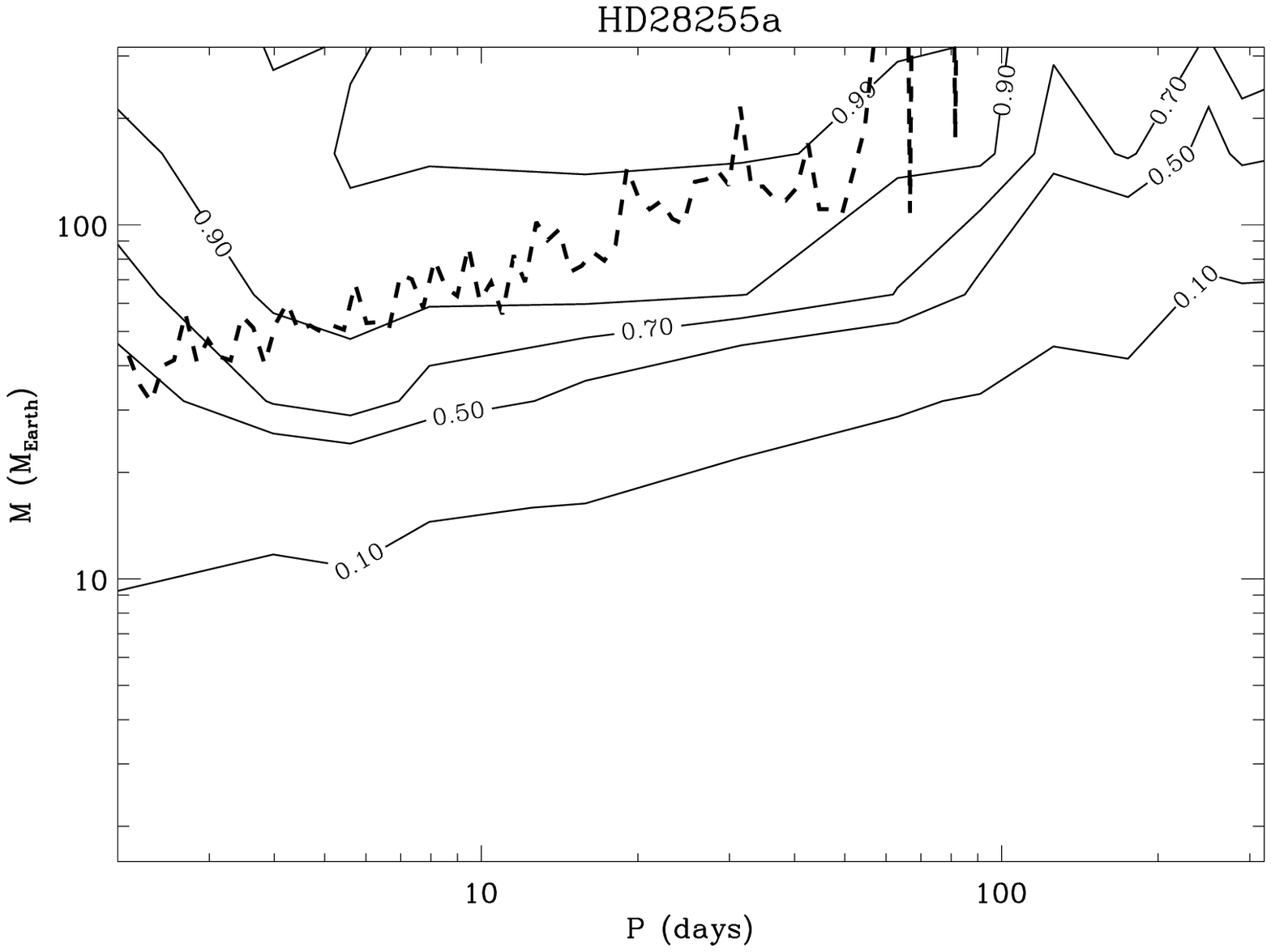}
\caption{Same as Figure~\ref{results1}, but for HD~26965 (left) and 
HD~28255A (right). }
\label{results5}
\end{figure}

\begin{figure}
\plottwo{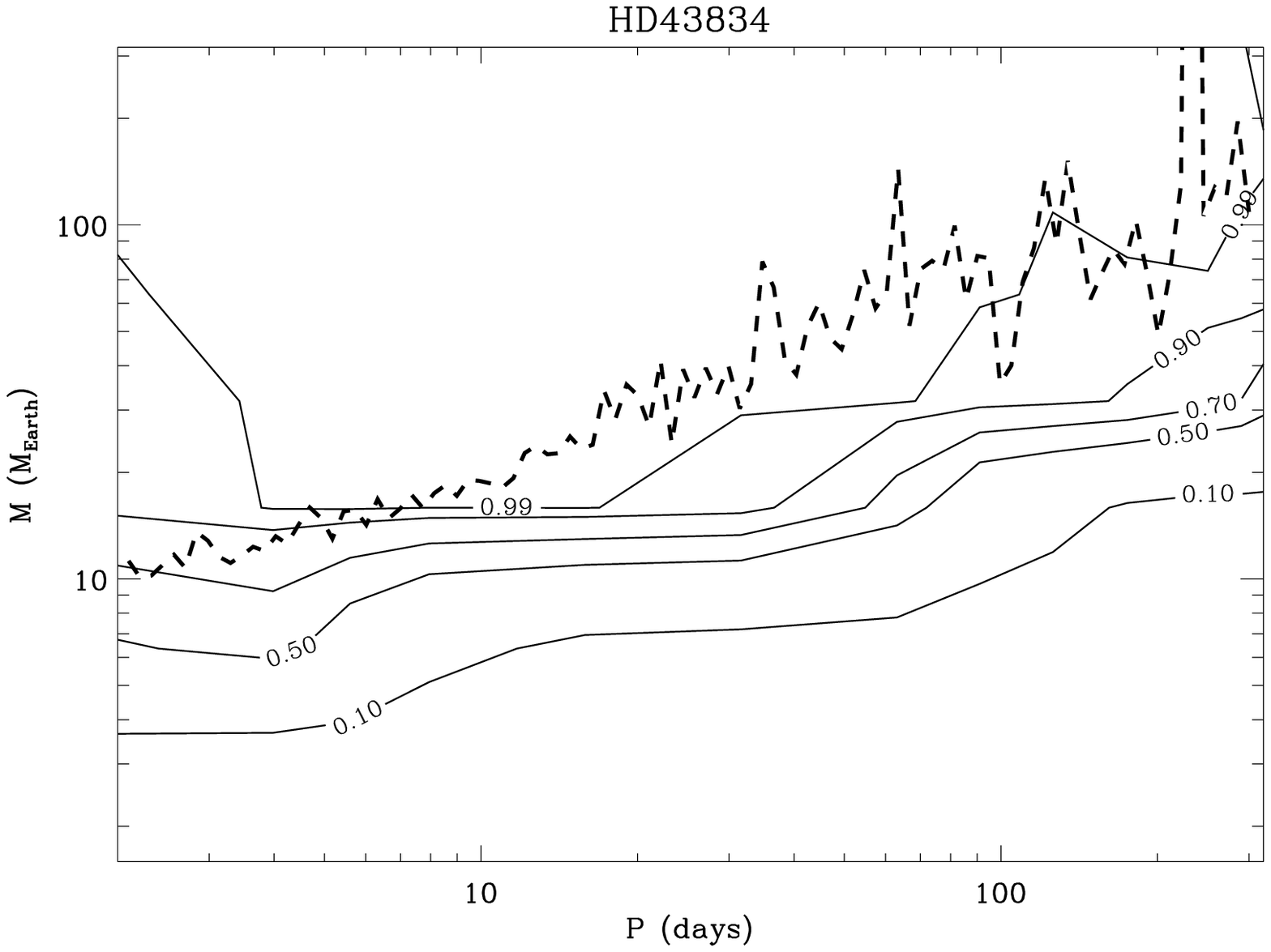}{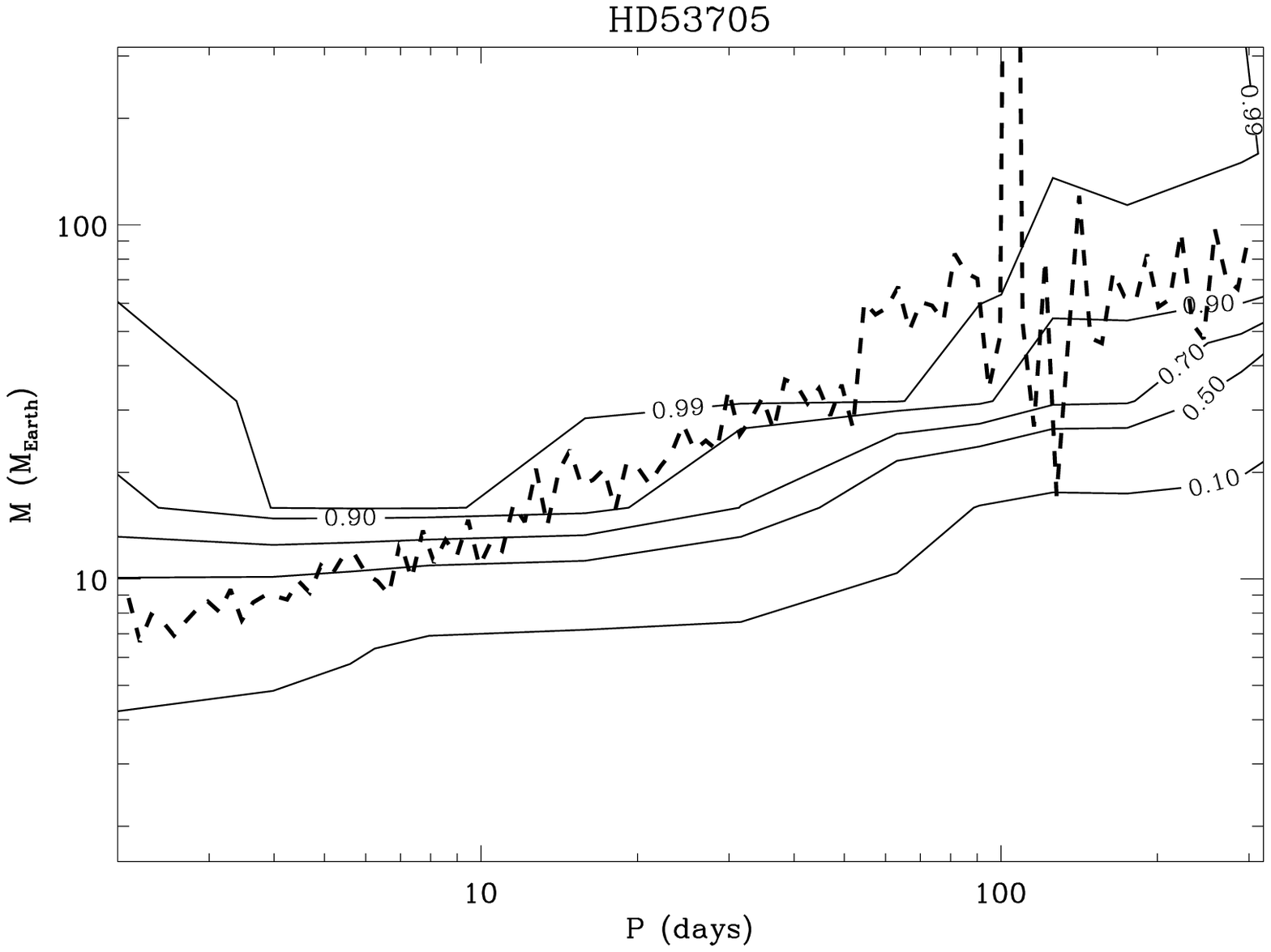}
\caption{Same as Figure~\ref{results1}, but for HD~43834 (left) and 
HD~53705 (right). }
\label{results6}
\end{figure}

\begin{figure}
\plottwo{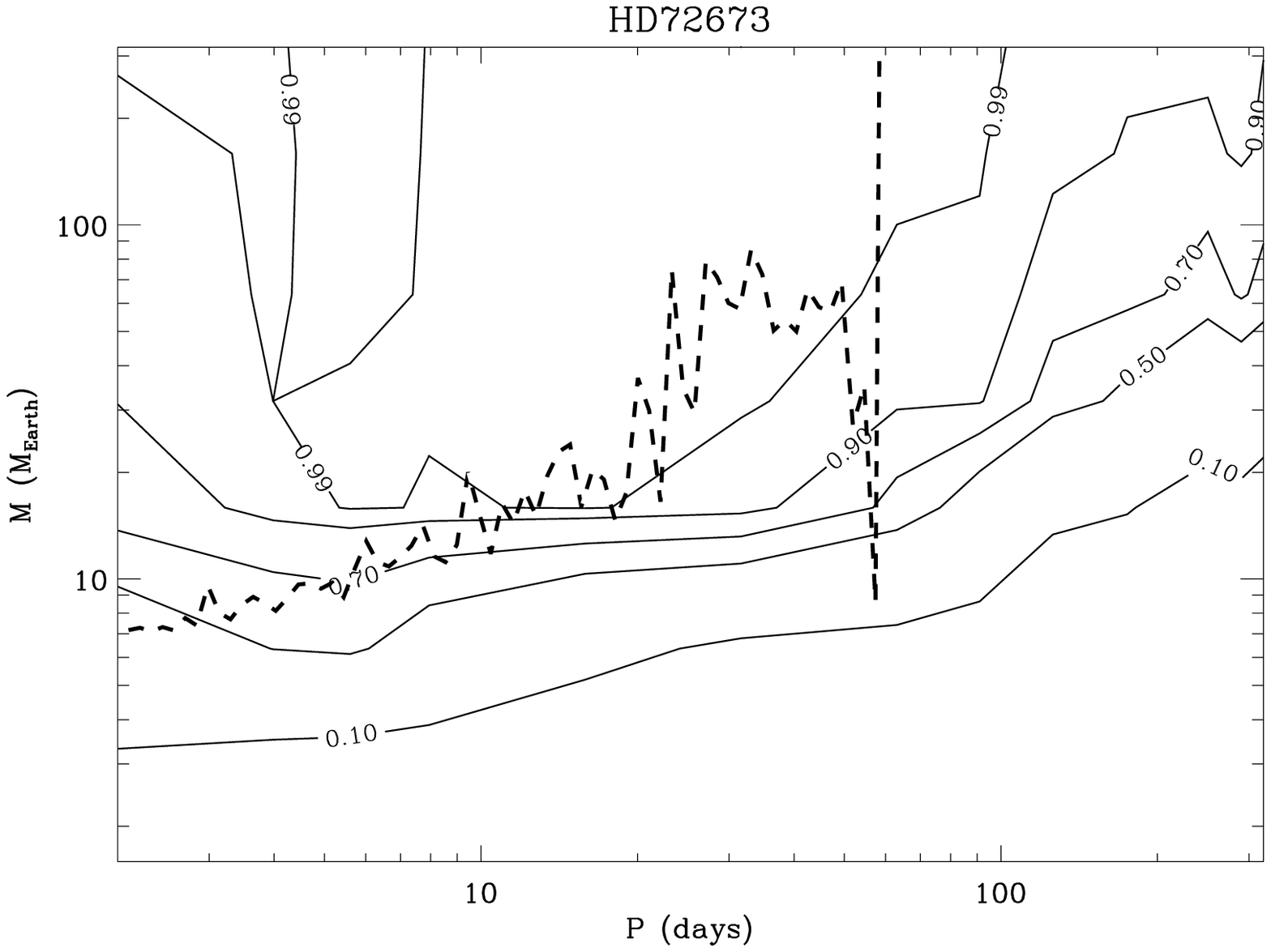}{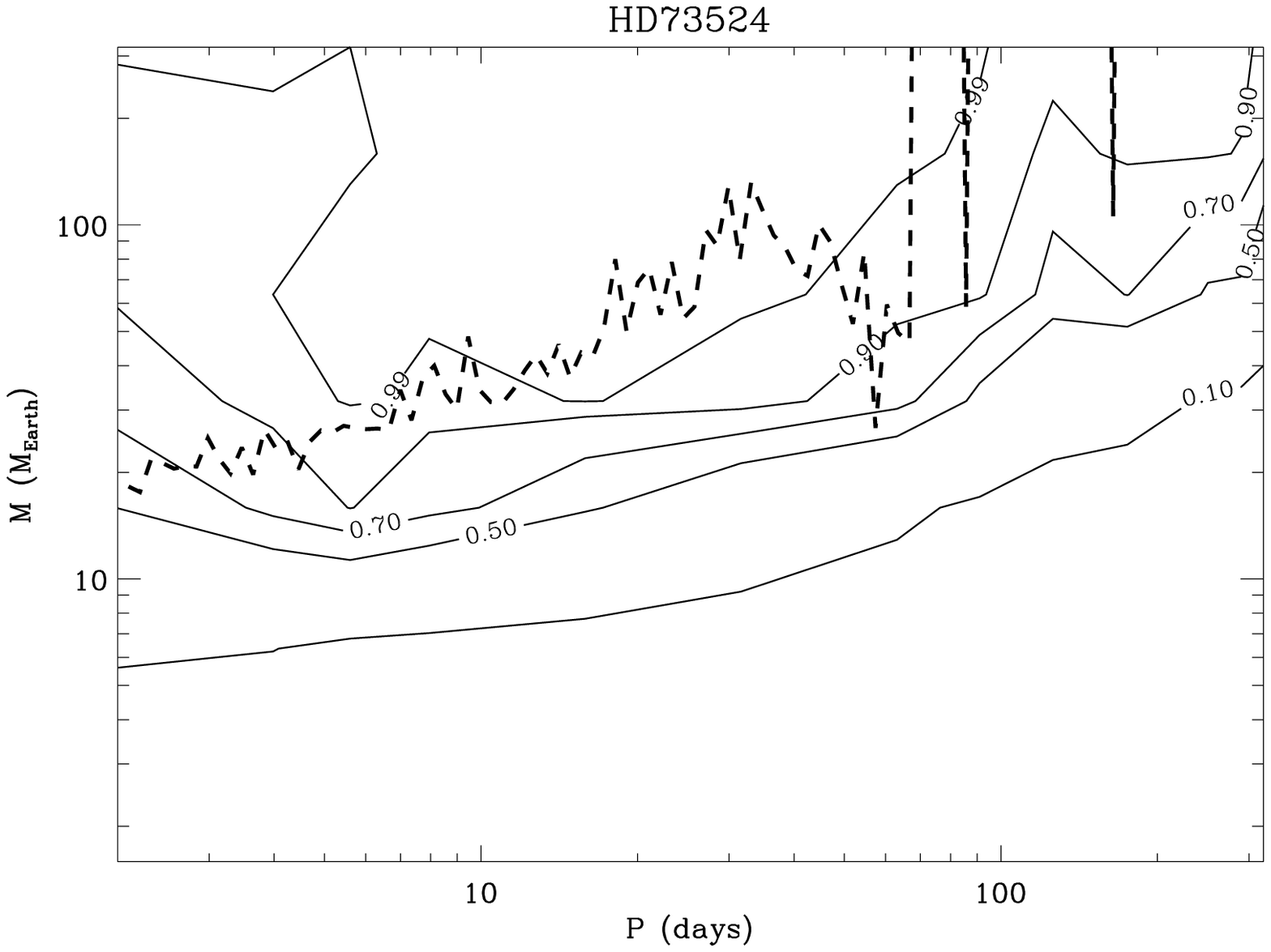}
\caption{Same as Figure~\ref{results1}, but for HD~72673 (left) and 
HD~73524 (right). }
\label{results7}
\end{figure}

\begin{figure}
\plottwo{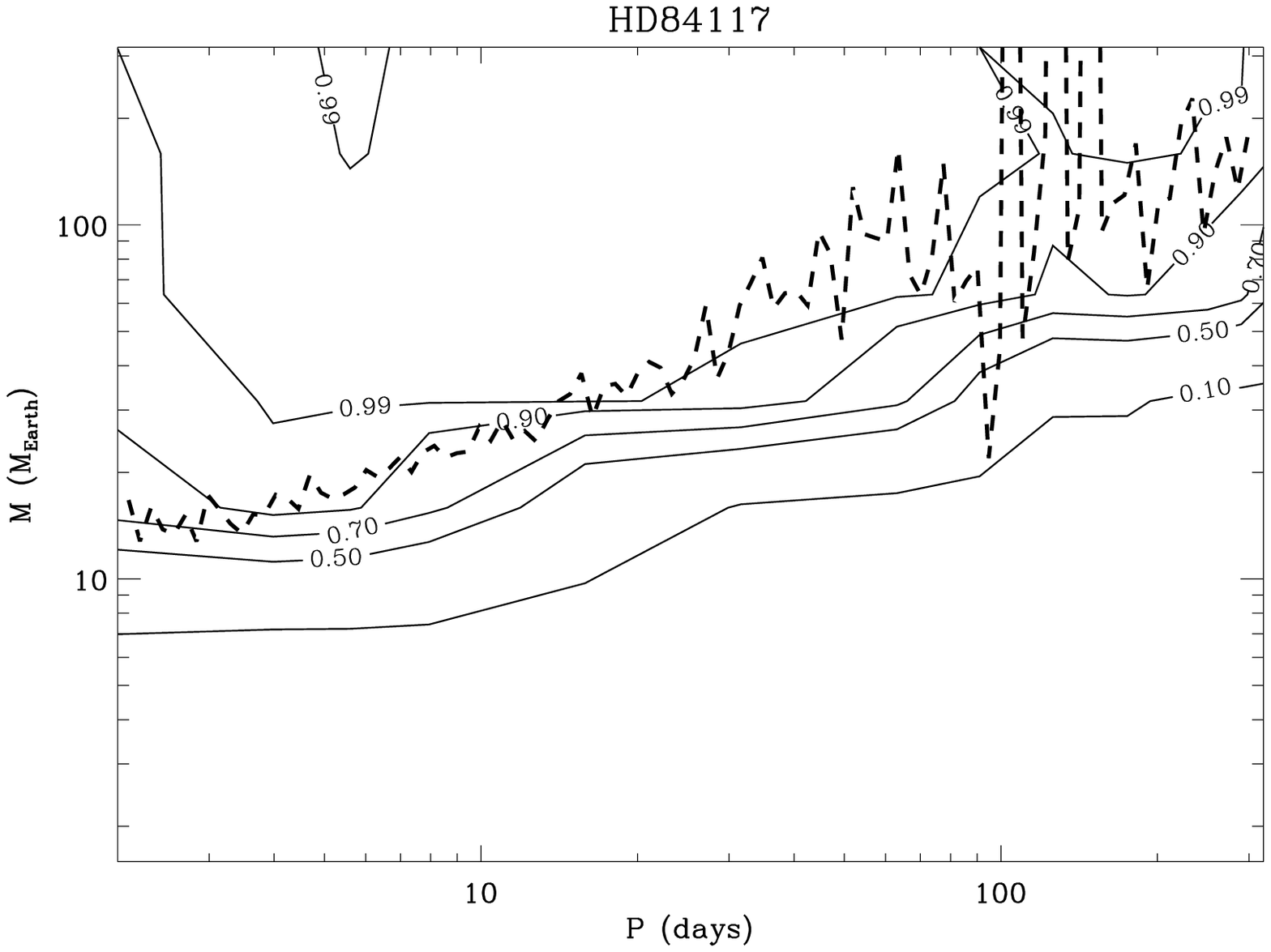}{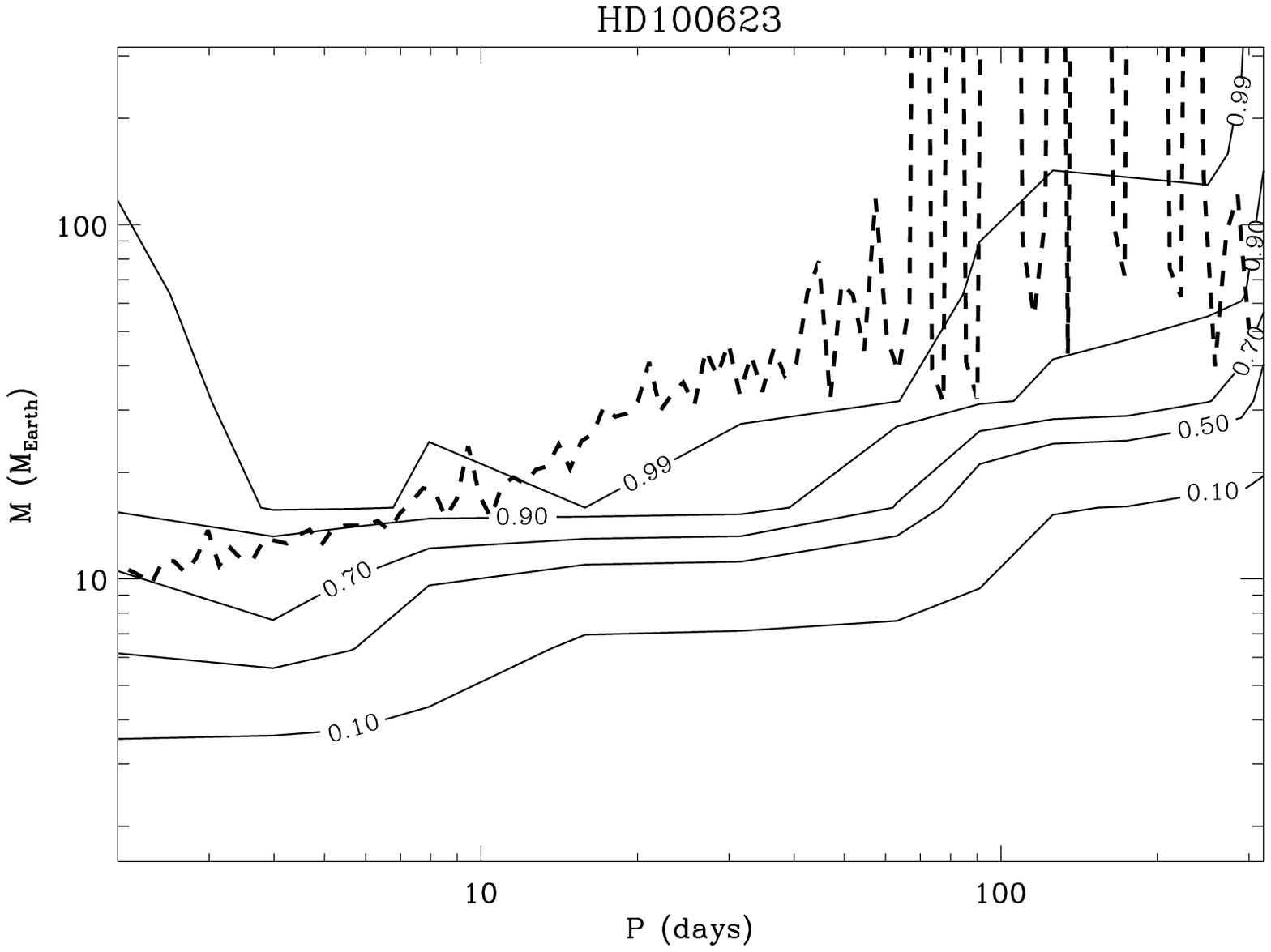}
\caption{Same as Figure~\ref{results1}, but for HD~84117 (left) and 
HD~100623 (right). }
\label{results8}
\end{figure}

\begin{figure}
\plottwo{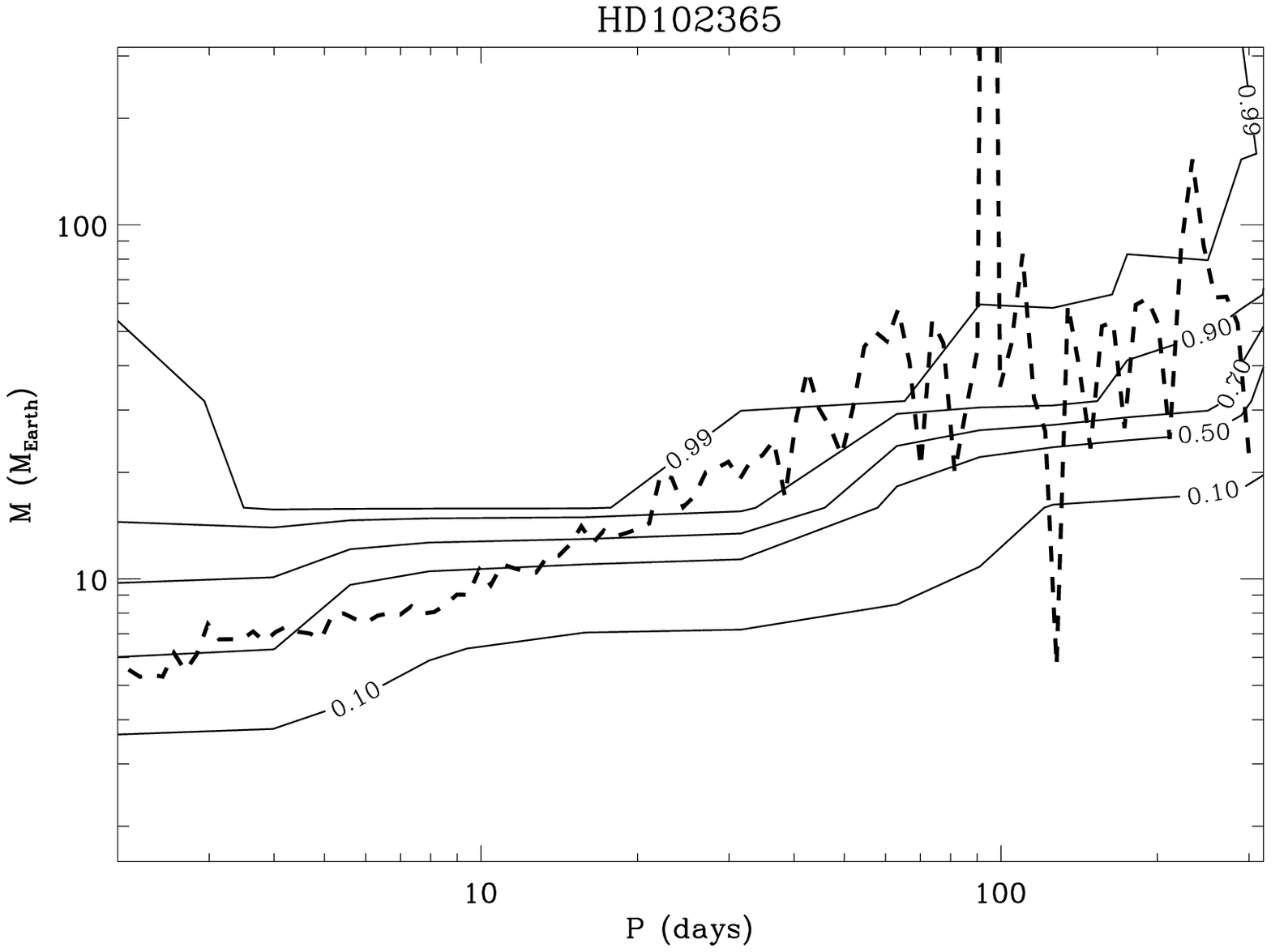}{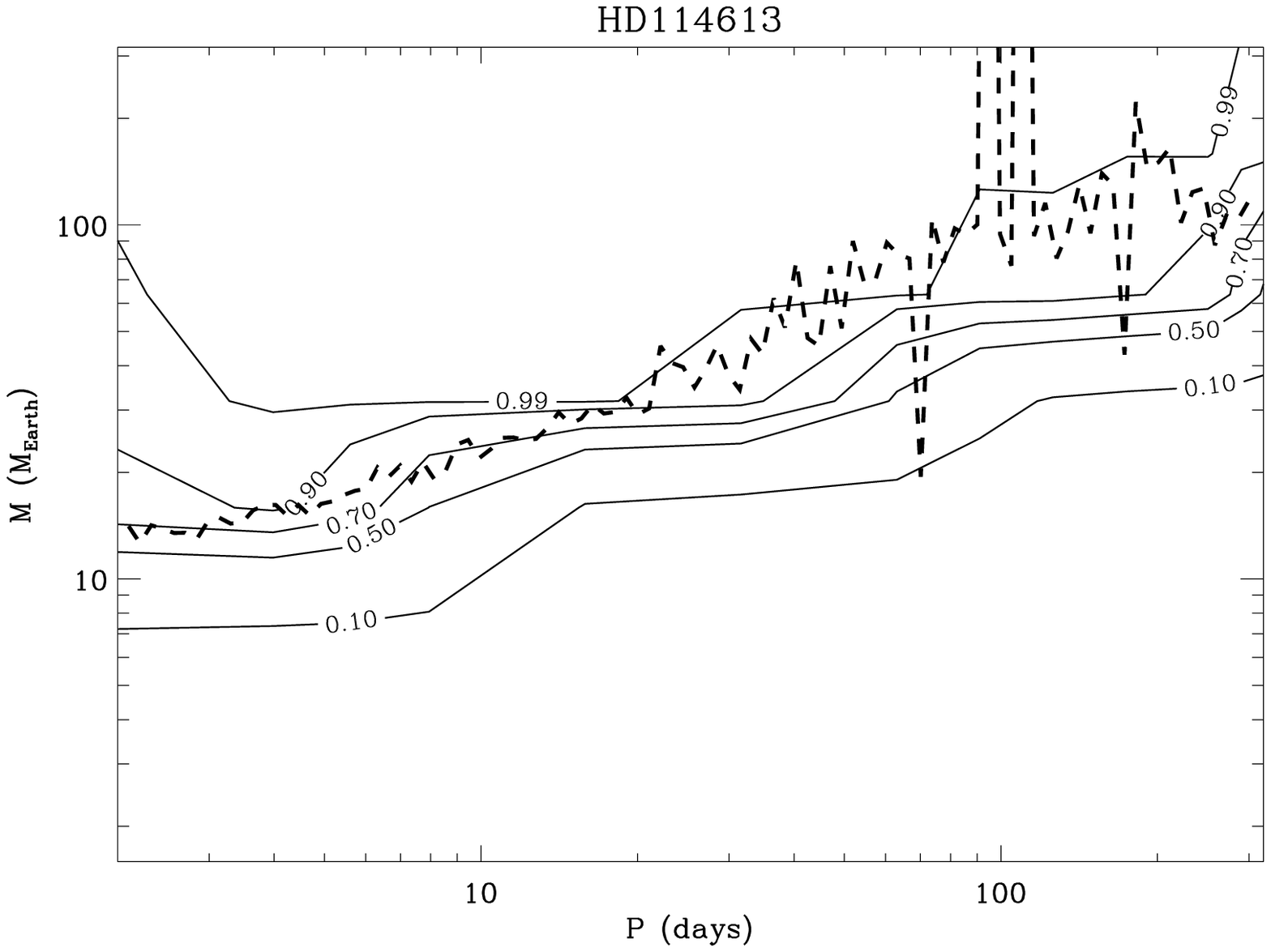}
\caption{Same as Figure~\ref{results1}, but for HD~102365 (left) and 
HD~114613 (right). }
\label{results9}
\end{figure}

\begin{figure}
\plottwo{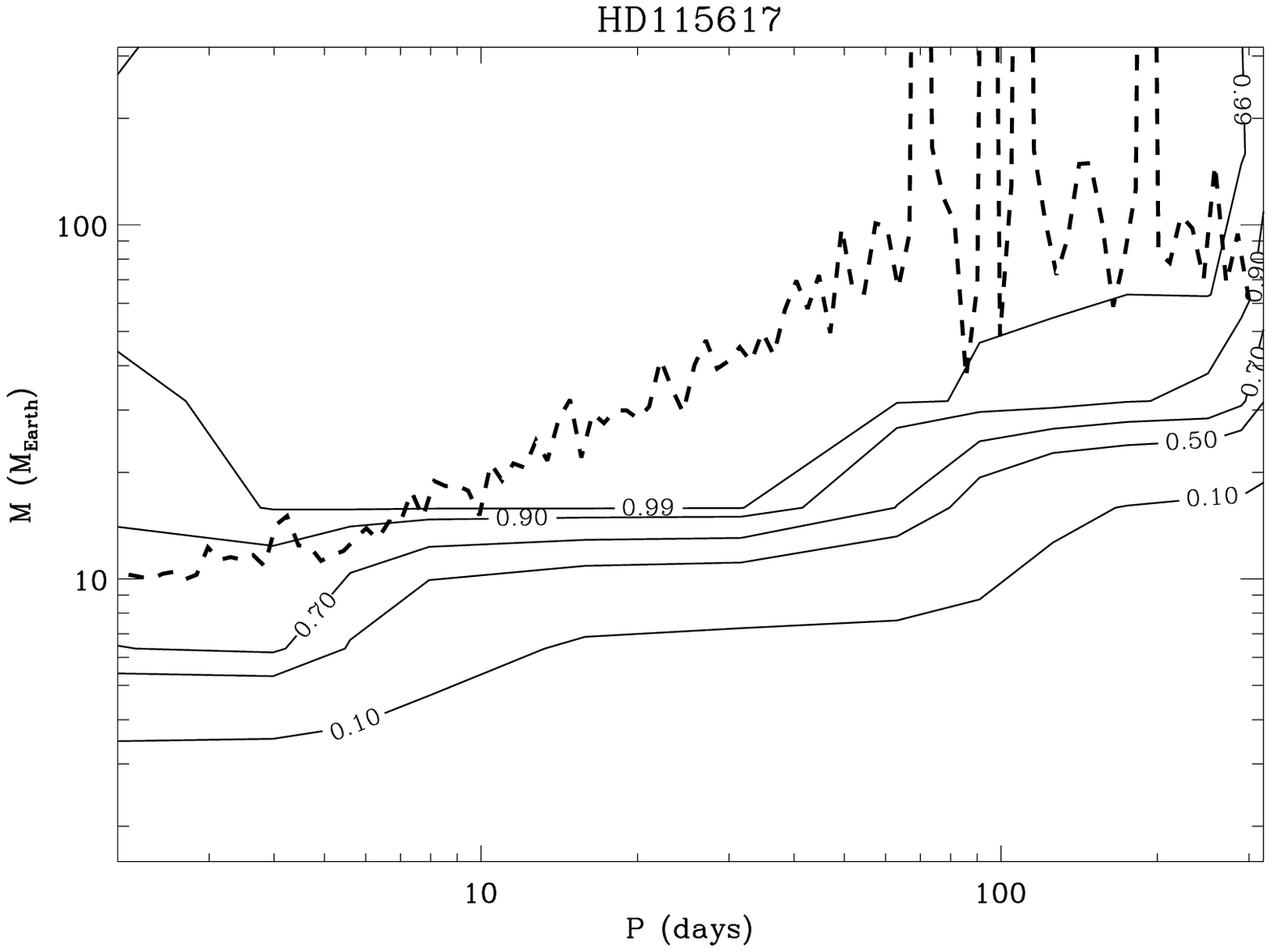}{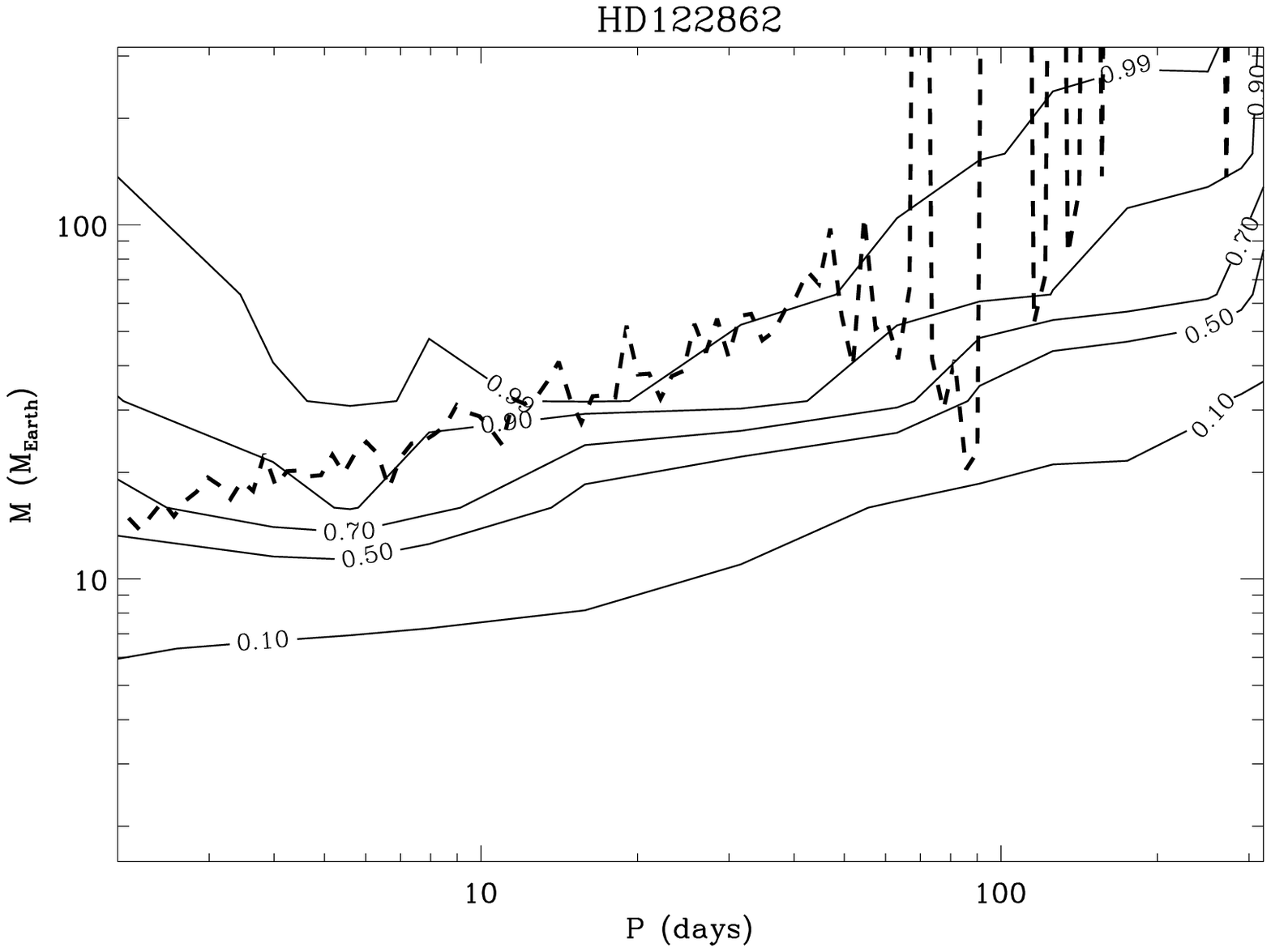}
\caption{Same as Figure~\ref{results1}, but for HD~115617 (left) and 
HD~122862 (right). }
\label{results10}
\end{figure}

\begin{figure}
\plottwo{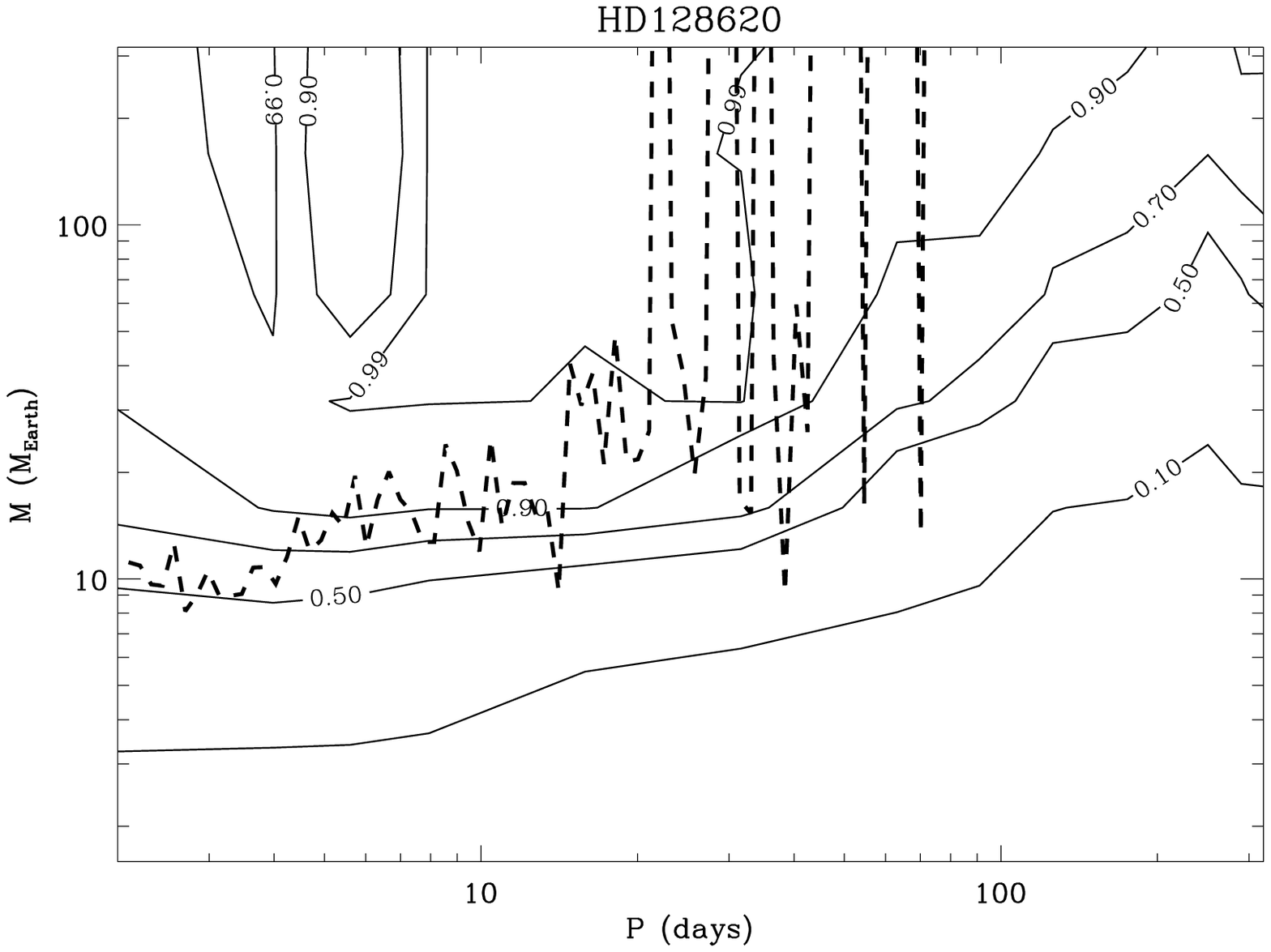}{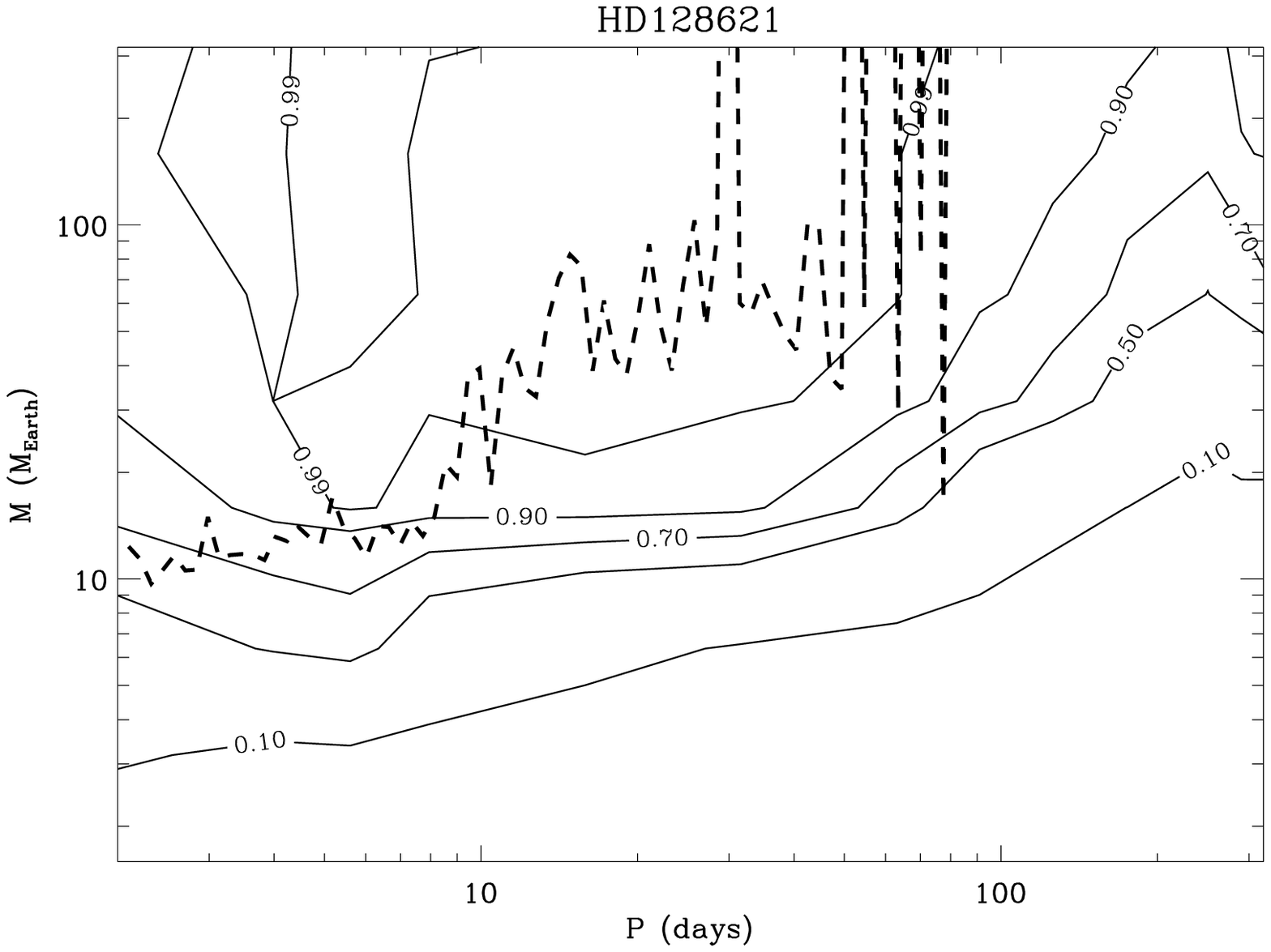}
\caption{Same as Figure~\ref{results1}, but for HD~128620 (left) and 
HD~128621 (right). }
\label{results11}
\end{figure}

\begin{figure}
\plottwo{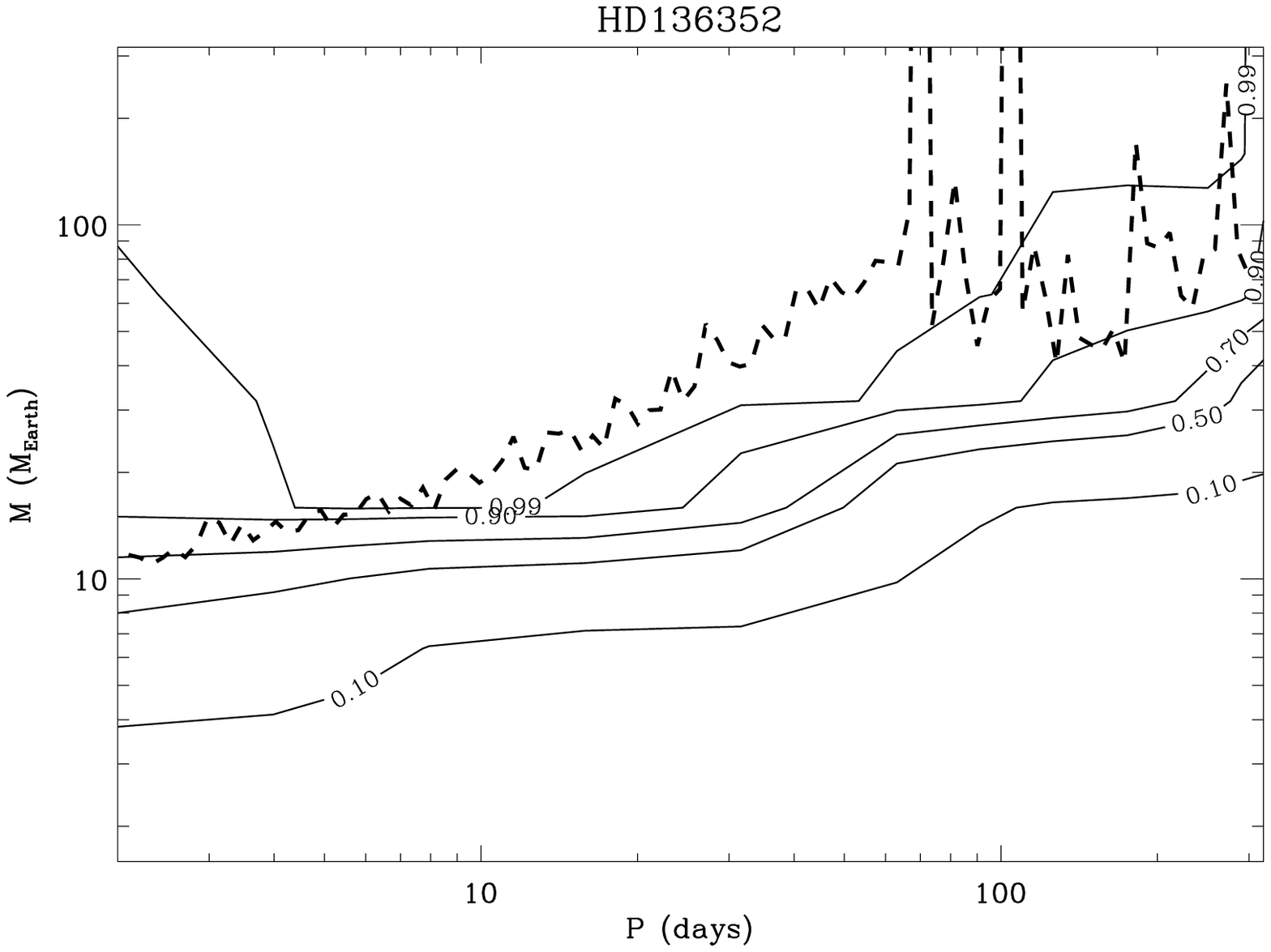}{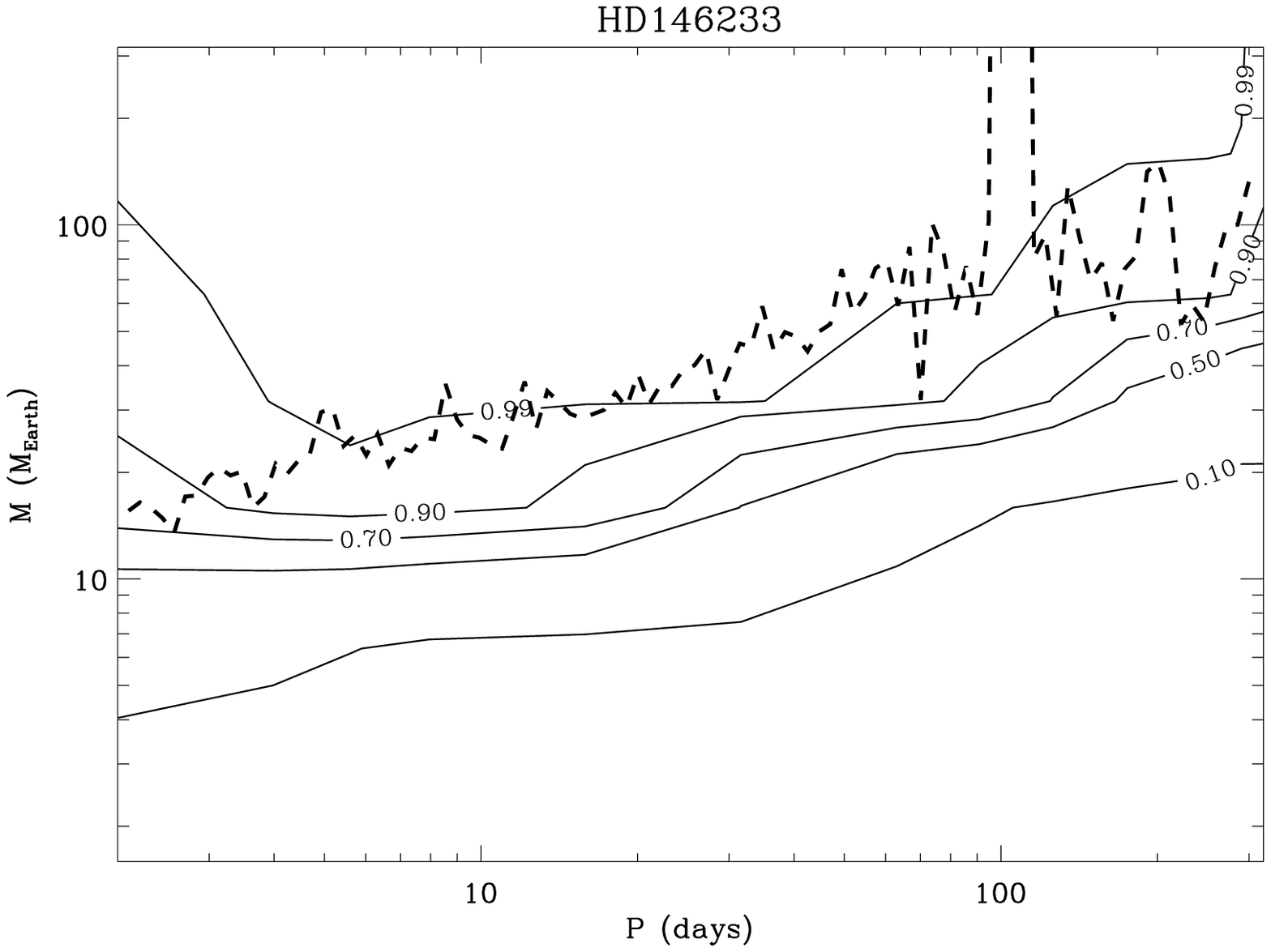}
\caption{Same as Figure~\ref{results1}, but for HD~136352 (left) and 
HD~146233 (right). }
\label{results12}
\end{figure}

For the purpose of this study, we define ``detectability'' as the 
fraction of simulated planets which were successfully recovered by the 
detection criteria employed by each method.  To obtain a ``big-picture'' 
view of the types of planets detectable with our current data set, we 
have averaged the results over all stars and all eccentricities.  
Figure~\ref{means} shows the detectabilities from Method~1; the contours 
indicate the fraction of simulated planets that were recovered.  The 
99\% detection limits obtained in this manner from Method~2 are 
overplotted as a dashed line.  Results for each individual star are 
shown in Figures~\ref{results1}--\ref{results12}. The detectabilities 
obtained in this work are highly dependent on the characteristics of the 
data set; HD~20794 (Figure~\ref{results4}) has a large number of 
observations with little scatter ($N=78$, rms=3.43 \ms), whereas 
HD~28255a (Figure~\ref{results5}) has few data points and considerable 
scatter ($N=37$, rms=7.57 \ms).  These results highlight the need to 
perform detection-limit computations on a star-by-star basis 
\citep{otoole09a} rather than ``whole of sample.'' 
Table~\ref{limitstable} summarises the results from both methods for 
each star.  We list the mean velocity amplitude $K$ (averaged over all 
periods 2-300 days) which was recovered at the 99\% and 50\% levels for 
each method.

\subsection{Comparing the Methods}

In general, the results from the two methods are in agreement, though 
for $P<10$ days, Method~1 gives mass limits higher than Method~2 by a 
factor of 2-4 (cf.~Fig.~\ref{means}).  The reason for this difference is 
that Method~1 incorporates the added criterion that the semi-amplitude 
$K$ of the recovered Keplerian orbit must satisfy $K/\sigma_{K}>2$.  
Removing this criterion results in an extremely high proportion of false 
positives (incorrect detections).  Method~2, with a periodogram 
recovery, has no such constraint on the amplitude, only on the period.  
Both are more conservative than a human would be in analysing an 
individual observed data set by hand.  On the other hand, they have a 
quantitative selection process which means that, given their detection 
criteria and a set of data, one can use these algorithms to draw robust 
conclusions about planetary frequency and orbital parameter 
distribution.

\begin{figure}
\plotone{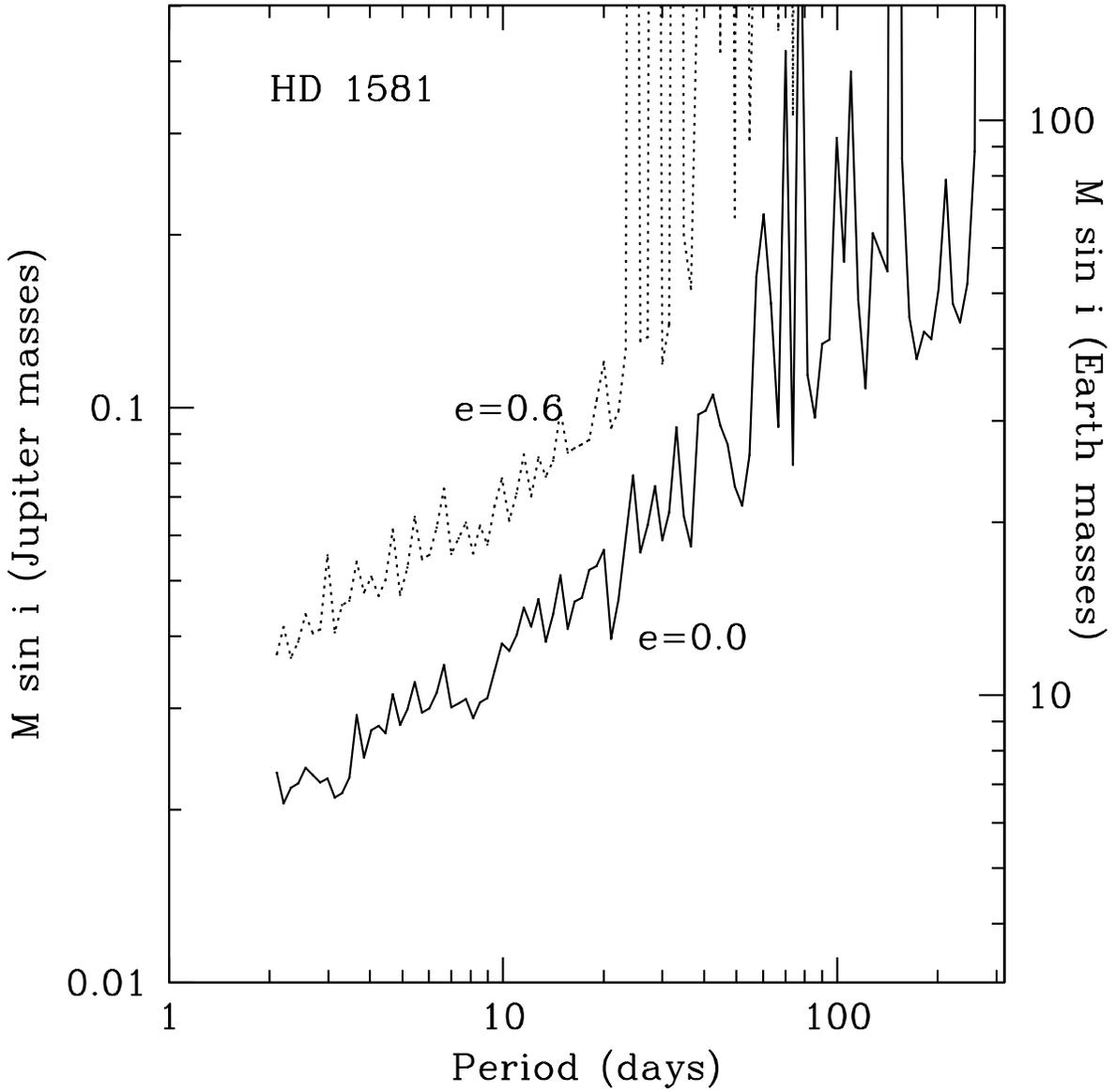}
\caption{Detection limits for HD~1581 (Method~2) for test signals with 
$e=0.0$ (solid line) and $e=0.6$ (dashed line).  At higher 
eccentricities, there arise some periods for which the test signal is 
never recovered at sufficient significance by the traditional 
Lomb-Scargle periodogram. } \label{blindspots}
\end{figure}

Both Methods~1 and 2 address the question of ``What planets could have 
been detected by these data?'' In any such detection-limit 
determination, the aim is to automate and perform thousands of times 
what a human investigator would reasonably do to detect a signal 
\citep{otoole09a}.  This process necessarily involves some sacrifices.  
For example, Method~2 attempts to detect an injected signal using the 
standard Lomb-Scargle periodogram, an extremely fast computation.  
However, the Lomb-Scargle periodogram is most efficient at detecting 
sinusoidal signals; when Keplerian orbits become moderately eccentric, 
their shapes can become significantly non-sinusoidal \citep{cumming08, 
otoole09a, wittenmyer09}.  To work around this issue, 
\citet{limitspaper} restricted the eccentricities of the injected 
planets to $e<0.6$, a range which includes about 90\% of known 
exoplanets.  The effects of eccentricity remain evident, however, in the 
``blind spots'' where Method~2 indicates that planets at certain periods 
are completely undetectable (Figure~\ref{blindspots}).  For each star, 
there are certain trial periods where the data sampling conspires with 
an injected eccentric signal, such that an ordinary Lomb-Scargle 
periodogram does \textit{not} exhibit a significant peak at that period.  
Method~2 would then reject the trial signal, even at a large amplitude 
($K_{max}=100$\ms), and these situations result in allegedly 
undetectable trial signals.  Method~1, however, using the 2DKLS 
periodogram which is much more effective at detecting eccentric signals, 
would not fall victim to this pathology.

The significance of a periodogram peak obtained by Method~2 is computed 
using the analytic formula given in \citet{hb86}.  However, the 
false-alarm probability (FAP) obtained by this formula tends to be 
smaller (i.e.~more significant) than that obtained by a bootstrap 
randomisation procedure \citep{kurster97}.  This is due to the fact that 
noise in radial-velocity data tends to be non-Gaussian.  Thus, a human 
would visually inspect the periodogram, then perform a bootstrap FAP 
calculation to obtain an accurate estimate of the believability of any 
peaks.  The bootstrap FAP method, however, requires hours to calculate, 
rather than the milliseconds it takes for the analytic method, and can 
therefore not be usefully employed in the analysis of tens of thousands 
of simulations.

Another necessary concession is the requirement that the recovered 
period be within 5\% of the injected period.  The purpose of this 
constraint is to exclude significant periodogram peaks at aliases of the 
true period.  Spurious periods were also often produced by the lunar 
cycle ($\sim$30 days), and the duration of the densely sampled ``Rocky 
Planet Search'' (48 days).  All previously published work using this 
method \citep{limitspaper, swiftpaper, wittenmyer09} has used this 5\% 
criterion.  This has the effect of preventing incorrect detections 
(false positives).  Method~1 does not have an explicit criterion of this 
sort, but the four criteria outlined in \S~2.1 serve to keep the rate of 
false positives below $\sim$1-2\%; a detailed discussion of false 
positives is given in \citet{otoole09a}.

To summarise, Method~1 seems to be most appropriate for significantly 
eccentric orbits because the simulated planets are fitted with a full 
Keplerian model.  Since a value of stellar velocity noise (jitter) is 
added to the simulated data in Method~1, that method is more effective 
when the radial-velocity noise is well-understood.  If the stellar 
jitter is not well-understood, it is more appropriate to use Method~2 
since the actual velocity data (which include all noise sources) are 
used.

% These differences may arise from (1) false negatives 
% from Method~1, and/or (2) false positives in Method~2.  False negatives 
% arise when a trial input signal is wrongly rejected, and false positives 
% occur when a recovered signal is incorrectly accepted \citep{otoole09a}.  

\section{Discussion}

\subsection{$M_p >$100 \Mearth }

\citet{jones03} and \citet{udry03} noted an emerging ``period valley'' 
between 10 and 100 days; the $\sim$250 additional planets discovered 
since then have continued this trend.  \citet{udry03} also showed that 
this feature is mainly attributable to the lack of massive (m sin $i >$ 
2\Mjup) planets in that period range.  Our results support this claim, 
since the 24-star sample contains three low-mass planets with periods 
between 10-100 days (HD~4308b, HD~16417b, HD~115617c), and the data can 
exclude planets with m sin $i >$ 0.5\Mjup\ at the 99\% level for all of 
this region (Figure~\ref{means}).  The deficit of higher-mass planets 
with $P<300$ days is readily apparent, suggesting that the period valley 
is not a selection effect; this is consistent with the conclusions 
reached by \citet{cumming08}.

\subsection{$M_p <$100 \Mearth }

\begin{figure}
\plotone{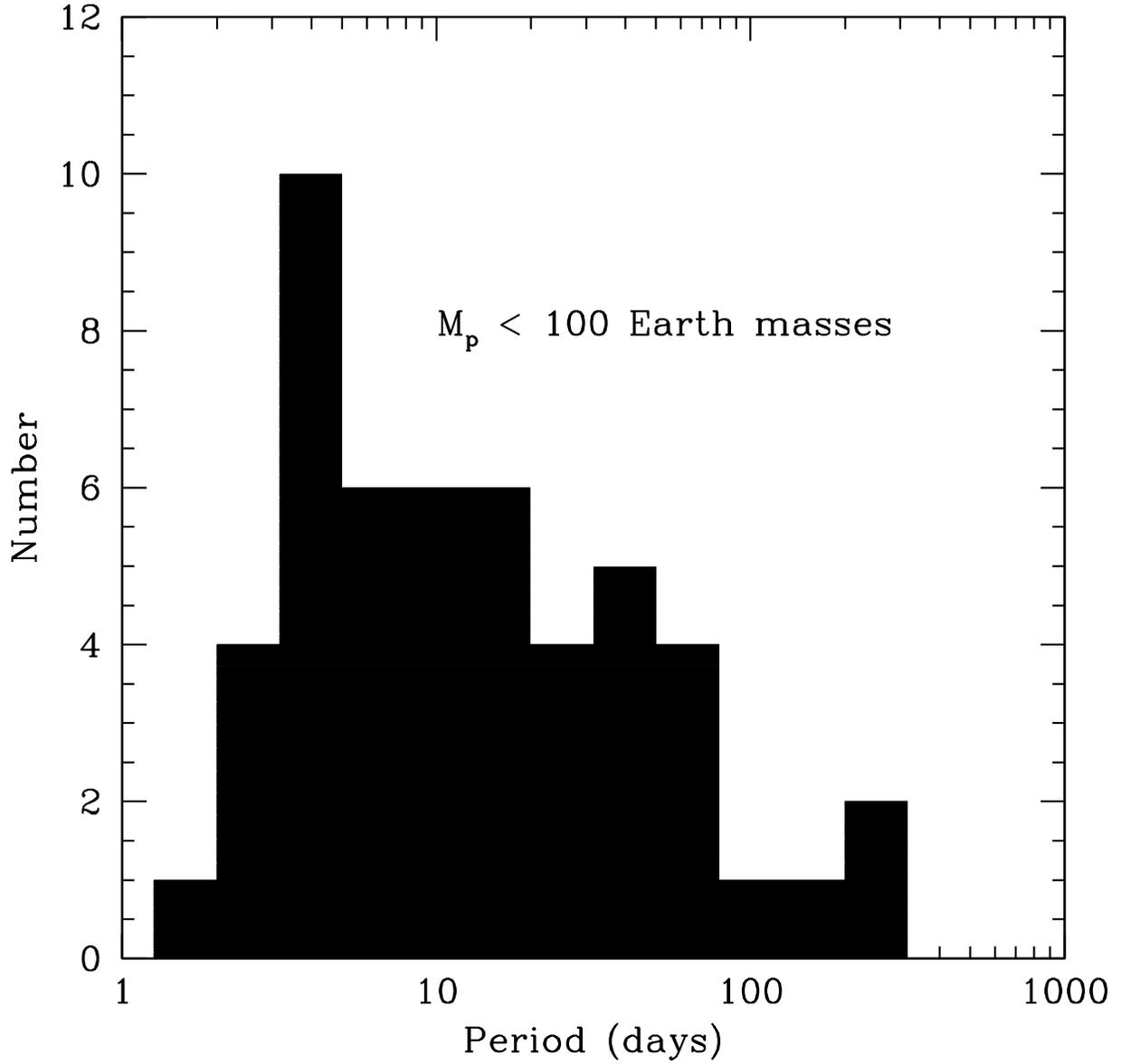}
\caption{Observed orbital period distribution of radial-velocity planets 
with m sin~$i <$100 \Mearth\ (N=50).  The fall-off at $P>5$ days arises 
from the twin selection effects of smaller velocity amplitudes and 
poorer temporal sampling, especially in the period range of 30-50 days. 
} \label{lowmass}
\end{figure}

At lower masses (10-100 \Mearth) however, the situation is far less 
clear.  This is an extremely interesting mass range, as core-accretion 
theory, and numerical simulations based on it, predict a ``planet 
desert'' to arise from rapid gas accretion by cores once they reach 
about 10\Mearth\ \citep{idalin04a, idalin08a, liu09}.  Recent 
simulations by \citet{liu09} showed a pile-up of planets at 0.2-0.3~AU 
($P\sim$30-50 days).  This is a mass range, then, which {\em should} 
show detectable features in the structure of the period distribution.  
However, this is also a mass range at which serious selection effects 
impact on the detectability of low-mass planets, and most especially in 
the critical 15-60d orbital period range.  This is why \citet{liu09} 
noted that the predicted pile-up had not yet been observed.  
Figure~\ref{lowmass} shows the period distribution of planets with $M_p 
<$100 \Mearth.  A steep fall-off in the distribution is evident at 
$P\gtsimeq$5 days.  This can be understood as arising from two selection 
effects.  First and most obviously, longer-period planets have weaker 
radial-velocity signals.  Second, planet search observing runs are 
almost exclusively scheduled on large telescopes during bright time, 
resulting in a large peaks in the window function at 28d.  The only 
means to mitigate this sampling problem is to observe through dark 
lunations.  This is a strategy that only the AAPS is currently employing 
in its ``Rocky Planet Search'' observing campaigns and which (as 
demonstrated by the discovery of 61~Vir~c, $P=38$d, $K=3.6$ \ms) has 
proven successful in finding low-mass planets in this period range.

When observing strategies are employed which mitigate selection effects 
in this period regime, we do indeed find low-mass planets (e.g.~Vogt et 
al.~2010, O'Toole et al.~2009b).  These discoveries indicate that the 
observed fall-off in the distribution of such planets 
(Figure~\ref{lowmass}) may -- contrary to what is seen at higher masses 
-- indeed merely be a result of observational biases.  If we look at the 
results of our simulations, they indicate that 10\% of planets with 
$K\sim$2.7 \ms\ could have been detected in our sample of 24 stars.  One 
such planet with an even smaller radial-velocity amplitude has atually 
been discovered in our sample (61~Vir~b, $K=2.2$ \ms), suggesting that 
at least a further nine more such planets could have been missed, and 
will remain below the detection threshold until additional data are 
obtained.

In addition, the solar-type dwarf stars typically observed by precision 
radial-velocity programs have typical rotation periods in the 20-50 day 
range.  Hence, emergent periodicities are not usually believed until 
sufficient data are in hand to be confident that the periodicity is 
planetary in origin rather than due to rotational modulation by 
starspots.  This selection bias suggests that low-mass planets with 
$P\sim$30-50 days may be present in extant radial-velocity data sets, 
but as yet not believed and unpublished.

\subsection{The underlying planetary mass function} 

The importance of planetary migration has been appreciated for some time 
\citep{goldreich80, lin86}.  The modelling work of \citet{trilling02} 
and \citet{armitage02} has provided evidence that observational data for 
exoplanets are consistent with a model where planets are formed at 
around 5~AU and then undergo migration.  This leads to a distribution of 
planetary semimajor axis $a$ that depends on log($a$).  Although this 
scenario provides a good match to observations (e.g.~Armitage 2007), 
migration rates at short orbital periods are expected to be different 
for different mass planets.  The simulations of \citet{rice08} indicate 
three regimes: (1) low-mass planets ($<$0.5\Mjup) migrate inwards and 
stall at approximately half the rotation period of the star, (2) 
high-mass planets (10\Mjup) should collide with the star, and (3) the 
intermediate-mass (1\Mjup) planet population should have an inner edge 
located at twice the Roche-limit \citep{rasio96}.  For the purposes of 
our models, we implement a distribution of planetary semimajor axis that 
depends on log($a$), referred to here as ''logarithmic migration.''

\begin{figure}
\includegraphics[height=5.5in,angle=90]{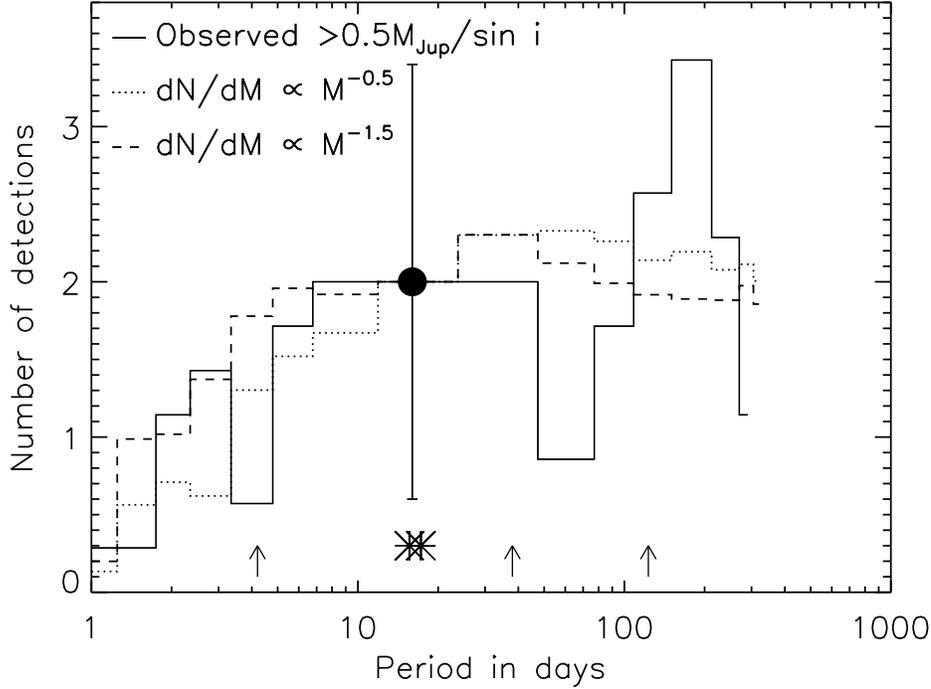}
\caption{ Solid-line histogram: Orbital period distribution of 
radial-velocity exoplanets with m sin $i >$ 0.5\Mjup\ (N=256); both 
simulation methods discussed in this work show that all of the planets 
this massive would have been detected in our sample, indicating that the 
lack of giant planets with $P<100$ days does not arise from selection 
biases.  Overplotted as dashed and dotted histograms are orbital period 
distributions that would be expected from our simulations, for a variety 
of assumed power-law mass functions.  Asterisks show the periods of 
announced exoplanets from our sample that are redetected by our 
automated selection criteria (\S~2), while arrows show the periods of 
the 3 planets detected orbiting 61\,Vir (using a combination of AAPS and 
Keck data), which are {\em not} redetected by our automated criteria 
with AAPS data alone. } \label{newfig7}
\end{figure}

Although our target star sample and number of detections is small, we 
can use the fact that it is well-characterised in an attempt to 
understand the features of the larger inhomogenous data set of radial 
velocity exoplanets.  Following the work of \citet{monster} and using 
our computed detectabilities and a simplistic logarithmic migration 
\citep{armitage07}, Figure~\ref{newfig7} shows the expected number of 
planets for different underlying mass functions as a function of 
semi-major axis for the case of equal numbers of exoplanets in each bin.  
The bold asterisks represent the planets detected from this sample 
(Table~\ref{planets}).  We find that the total expected number of 
detections for different assumed mass functions observed with our sample 
varies by a factor of about two for the mass functions shown, with all 
giving a smooth fall-off in the number of detections toward longer 
periods.  Thus, although the peak of the histogram serves to 
approximately characterise our detection of two exoplanets in the 10-20 
day period bin and is consistent with those below the threshold, the 
clear rise at longer periods in the observed distribution in 
Figure~\ref{newfig7} is not reproduced.  However, it must be noted that 
the sample of all radial velocity exoplanets shown in 
Figure~\ref{newfig7} is highly inhomogeneous and represents an observed 
sample of a higher mass range.

\begin{figure}
\includegraphics[height=5.5in,angle=90]{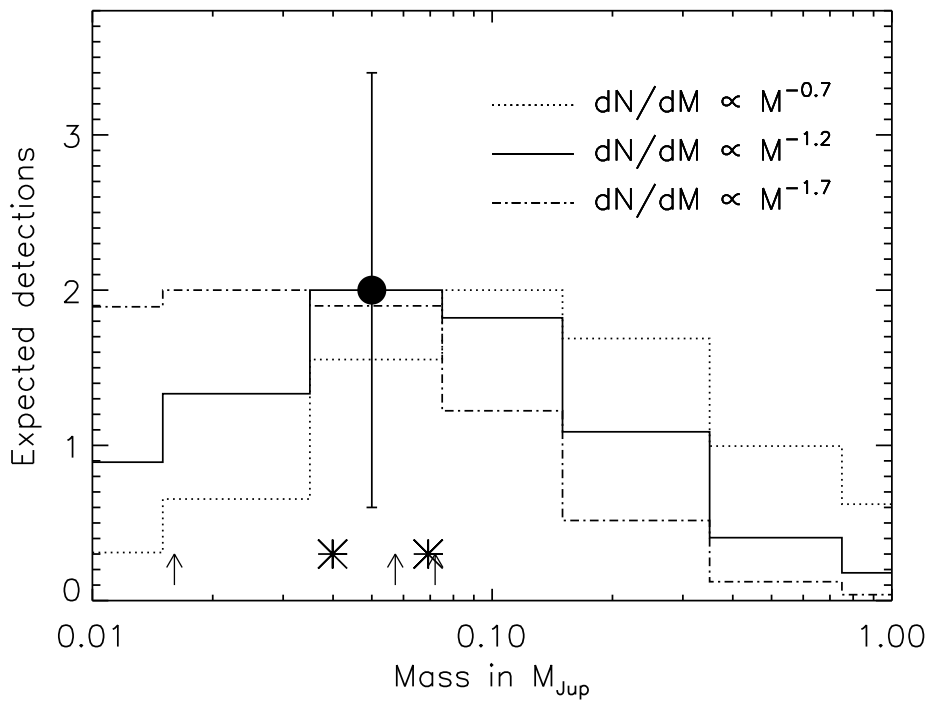}
% \plotone{f18.ps}
\caption{Expected number of exoplanets from our sample as a function of 
mass, for different mass functions including logarithmic migration.  
Asterisks and arrows show the mass of announced exoplanets from our 
sample, as for Figure~\ref{newfig7}.  The heavy filled circle with 
uncertainties shows the number density represented by our automated 
system detections in the 0.05\Mjup\ bin, while arrows show the periods 
of the 3 planets detected orbiting 61\,Vir (using a combination of AAPS 
and Keck data), which are {\em not} redetected by our automated criteria 
with AAPS data alone. }
\label{rob4}
\end{figure}

Although Figure~\ref{newfig7} indicates that the distribution in period 
space is relatively insensitive to different mass functions, we find 
greater sensitivity when detectability is plotted as a function of mass. 
Following \citet{monster}, we show in Figure~\ref{rob4} the expected 
detections for our sample as a function of mass for a variety of assumed 
mass functions.  We find that the peak in the expected number of {\em 
observed} exoplanets moves for different mass functions.  
$\alpha$=$-1.7$ leads to a peak in the number of detected planets around 
0.02\,\Mjup\ (6.3\,\Mearth), the $\alpha$=$-1.2$ expected detections 
peak at around 0.05\,\Mjup\ (16\,\Mearth) while the flatter 
$\alpha$=$-0.7$ mass function produces an expected detection peak at 
around 0.10\,\Mjup\ (32\,\Mearth).

Given that we have found two exoplanet signals in the 16-25\,\Mearth\ 
minimum mass range (corresponding to the 0.05\,\Mjup\ or 16\,\Mearth\ 
bin in Figure~\ref{rob4} we may consider these as providing rough limits 
on the underlying mass function as determined by how changes in the mass 
function cause the peak of detections to move toward and away from the 
0.05\,\Mjup\ (16\,\Mearth) bin. We find the peak moves from the 0.02 to 
the 0.05\,\Mjup\ bin for $\alpha$ values shallower than $\alpha$=$-1.6$ 
and from 0.05\,\Mjup\ to 0.1 \Mjup\ at $\alpha$=$-1.1$. It is noteworthy 
that the peak mass for different mass functions is sensitive to 
migration. If no migration is employed then the mass function peak moves 
away from the 0.05\,\Mjup\ (16\,\Mearth) bin when 0.6 $\leq$ $\alpha$ 
$\leq$ 1.2.  Based on Poisson-counting statistics of two detections our 
simulations indicate a normalisation of 0.17$\pm$0.12 (corresponding to 
the F value presented in \citet{monster}).  Our estimate of the 
underlying mass function shown in Figure~\ref{rob4} is consistent with 
the first attempt presented in \citet{monster}.  That is, the additional 
planet detections in this sample and the comprehensive detection limits 
presented here serve to strengthen the case for an underlying planet 
mass function of the form $dN/dM\propto M^{-1.0}$.  Given that we 
robustly do not detect more massive planets in this sample, flatter mass 
functions are not favoured.

\section{Conclusions}

We have shown, using two independent simulation algorithms, that recent 
data from the AAPS are capable of placing meaningful limits on the 
population of potentially terrestrial, sub-Neptune-mass planets orbiting 
nearby stars.  As we are able to have detected all planets with at least 
a Saturn mass in the period range $P<300$ days, we conclude that the 
observed lack of massive planets ($M_p >$100 \Mearth) with periods 
10-100 days is real and not due to selection effects.  However, we 
conclude that the shortage of lower-mass planets ($M_p <$100 \Mearth) 
predicted by core-accretion theory in this period range may arise from 
observational selection biases against detecting such planets.

%--------------------------------------------------------------------
\acknowledgements

We gratefully acknowledge the UK and Australian government support of 
the Anglo-Australian Telescope through their PPARC, STFC and DIISR 
funding; STFC grant PP/C000552/1, ARC Grant DP0774000 and travel support 
from the Anglo-Australian Observatory.  RW is supported by a UNSW 
Vice-Chancellor's Fellowship.

Exoplanet data were obtained from the Extrasolar Planets Encyclopedia 
(exoplanet.eu) maintained by Jean Schneider.  This research has made use 
of NASA's Astrophysics Data System (ADS), and the SIMBAD database, 
operated at CDS, Strasbourg, France.

%--------------------------------------------------------------------

%----------------------------------------------------------

\begin{deluxetable}{lrrr}
\tabletypesize{\scriptsize}
\tablecolumns{4}
\tablewidth{0pt}
\tablecaption{Summary of Radial-Velocity Data }
\tablehead{
\colhead{Star} & \colhead{$N$} & \colhead{RMS} & \colhead{$<\sigma>$} \\
\colhead{} & \colhead{} & \colhead{(\ms)} & \colhead{(\ms)}
 }
\startdata
\label{rvdata1}
HD 1581 & 58 & 2.81 & 0.95 \\
HD 4308 & 77 & 2.92\tablenotemark{a} & 1.09 \\
HD 10360 & 31 & 3.77\tablenotemark{b} & 0.90 \\
HD 10361 & 31 & 3.20\tablenotemark{b} & 0.84 \\ 
HD 10700 & 95 & 2.65 & 0.86 \\ 
HD 16417 & 55 & 2.66\tablenotemark{a} & 0.86 \\ 
HD 20794 & 78 & 3.43 & 0.81 \\ 
HD 23249 & 64 & 3.16 & 0.56 \\ 
HD 26965 & 55 & 4.80 & 0.76 \\ 
HD 28255A & 37 & 7.57 & 1.18 \\  % latest points are way up 
HD 43834 & 78 & 4.27 & 0.80 \\ 
HD 53705 & 80 & 3.09 & 1.21 \\ 
HD 72673 & 50 & 2.66 & 1.06 \\ 
HD 73524 & 44 & 4.22 & 1.28 \\ 
HD 84117 & 83 & 5.00 & 1.33 \\ 
HD 100623 & 62 & 4.08 & 1.00 \\ 
HD 102365 & 97 & 2.49 & 0.89 \\ 
HD 114613 & 106 & 4.25 & 0.78 \\ 
HD 115617 & 94 & 4.29 & 0.74 \\ 
HD 122862 & 49 & 3.81 & 1.27 \\ 
HD 128620 & 33 & 1.94\tablenotemark{c} & 0.40 \\ 
HD 128621 & 42 & 2.76\tablenotemark{c} & 0.56 \\ 
HD 136352 & 78 & 4.34 & 1.07 \\ 
HD 146233 & 48 & 3.91 & 1.02 \\ 
\enddata
\tablenotetext{a}{After removal of known planet's orbit.}
\tablenotetext{b}{Residuals of linear fit.}
\tablenotetext{c}{Residuals of quadratic fit.}
\end{deluxetable}

%----------------------------------------------------------
\begin{deluxetable}{lr@{$\pm$}lr@{$\pm$}lr@{$\pm$}lr@{$\pm$}lr@{$\pm$}lr@{$\pm$}
lr@{$\pm$}ll}
\rotate
\tabletypesize{\scriptsize}
\tablecolumns{9}
\tablewidth{0pt}
\tablecaption{Planets From This Sample }
\tablehead{
\colhead{Planet} & \multicolumn{2}{c}{Period} & \multicolumn{2}{c}{$T_0$}
&
\multicolumn{2}{c}{$e$} & \multicolumn{2}{c}{$\omega$} &
\multicolumn{2}{c}{K } & \multicolumn{2}{c}{M sin $i$ } &
\multicolumn{2}{c}{$a$ } & \colhead{Reference} \\
\colhead{} & \multicolumn{2}{c}{(days)} & \multicolumn{2}{c}{(JD-2400000)}
&
\multicolumn{2}{c}{} &
\multicolumn{2}{c}{(degrees)} & \multicolumn{2}{c}{(\ms)} &
\multicolumn{2}{c}{(\Mearth)} & \multicolumn{2}{c}{(AU)} & \colhead{}
 }
\startdata
\label{planets}
HD 16417 b & 17.24 & 0.01 & 50099.7 & 3.3 & 0.20 & 0.09 & 77 & 26 &
5.0 & 0.4 & 22.1 & 2.0 & 0.14 & 0.01 & \citet{16417paper} \\
HD 4308 b & 15.609 & 0.007 & 50108.5 & 1.9 & 0.27 & 0.12 & 210 & 21 &
3.6 & 0.3 & 13.0 & 1.4 & 0.118 & 0.009 & \citet{monster} \\
HD 115617 b & 4.2150 & 0.0006 & 53369.166 & (fixed) & 0.12 & 0.11 & 105 & 54 &
2.12 & 0.23 & 5.1 & 0.5 & 0.050201 & 0.000005 & \citet{61vir} \\
HD 115617 c & 38.021 & 0.034 & 53369.166 & (fixed) & 0.14 & 0.06 & 341 & 38 &
3.62 & 0.23 & 18.2 & 1.1 & 0.2175 & 0.0001 & \citet{61vir} \\
HD 115617 d & 123.01 & 0.55 & 53369.166 & (fixed) & 0.35 & 0.09 & 314 & 20 &
3.25 & 0.39 & 22.9 & 2.6 & 0.476 & 0.001 & \citet{61vir} \\
\enddata
\end{deluxetable}
%----------------------------------------------------------

%  used planet-fit method for 16417,4308
\begin{deluxetable}{lrrrr}
\tabletypesize{\scriptsize}
\tablecolumns{5}
\tablewidth{0pt}
\tablecaption{Summary of Detection Limits at e=0.0 }
\tablehead{
\colhead{} & \multicolumn{4}{c}{Mean $K$ velocity recovered (\ms)} \\
\colhead{} & \colhead{99\% recovery} & \colhead{99\% recovery} & 
\colhead{50\% recovery} & \colhead{50\% recovery} \\
\colhead{Star} & \colhead{Method 1} & \colhead{Method 2} & 
\colhead{Method 1} & \colhead{Method 2}
 }
\startdata
\label{limitstable}
HD 1581 & 12.6$\pm$1.1 & 4.8$\pm$2.2 & 6.0$\pm$0.5 & 2.8$\pm$0.6 \\
HD 4308 & 5.9$\pm$0.1 & 4.7$\pm$2.0 & 3.2$\pm$0.1 & 2.7$\pm$0.8 \\
HD 10360 & 11.9$\pm$0.3 & 9.6$\pm$1.8 & 3.4$\pm$0.1 & 7.0$\pm$3.1 \\
HD 10361 & 14.2$\pm$0.3 & 7.9$\pm$2.0 & 3.4$\pm$0.1 & 5.6$\pm$2.5 \\
HD 10700 & 5.0$\pm$0.2 & 3.5$\pm$0.9 & 2.8$\pm$0.1 & 2.4$\pm$0.4 \\
HD 16417 & 5.8$\pm$0.3 & 5.9$\pm$2.5 & 3.1$\pm$0.2 & 3.4$\pm$0.9 \\
HD 20794 & 4.8$\pm$0.1 & 4.3$\pm$1.0 & 2.4$\pm$0.1 & 2.7$\pm$0.5 \\
HD 23249 & 12.7$\pm$3.0 & 4.6$\pm$1.9 & 6.2$\pm$1.4 & 2.8$\pm$0.7 \\
HD 26965 & 5.3$\pm$0.6 & 8.3$\pm$3.7 & 2.5$\pm$0.3 & 6.2$\pm$1.8 \\
HD 28255A & 34.2$\pm$10.2 & 14.1$\pm$3.2 & 8.8$\pm$2.4 & 11.4$\pm$5.5 \\
HD 43834 & 4.6$\pm$0.4 & 6.3$\pm$2.1 & 2.5$\pm$0.2 & 4.0$\pm$0.8 \\
HD 53705 & 5.7$\pm$0.3 & 5.0$\pm$2.6 & 3.1$\pm$0.2 & 3.0$\pm$0.9 \\
HD 72673 & 14.5$\pm$0.5 & 5.4$\pm$2.8 & 3.3$\pm$0.1 & 4.7$\pm$4.3 \\
HD 73524 & 14.4$\pm$1.2 & 8.6$\pm$3.8 & 4.1$\pm$0.3 & 6.8$\pm$4.0 \\
HD 84117 & 7.9$\pm$0.1 & 7.9$\pm$3.4 & 4.1$\pm$0.1 & 5.1$\pm$1.5 \\
HD 100623 & 6.5$\pm$0.8 & 7.2$\pm$3.5 & 2.9$\pm$0.4 & 4.8$\pm$1.5 \\
HD 102365 & 5.6$\pm$0.1 & 3.9$\pm$1.6 & 2.9$\pm$0.1 & 2.5$\pm$0.6 \\
HD 114613 & 8.3$\pm$0.1 & 6.7$\pm$2.9 & 4.3$\pm$0.1 & 4.5$\pm$1.0 \\
HD 115617 & 4.0$\pm$0.2 & 7.3$\pm$3.2 & 2.4$\pm$0.1 & 4.8$\pm$1.4 \\
HD 122862 & 8.9$\pm$1.1 & 6.4$\pm$2.0 & 4.0$\pm$0.5 & 4.8$\pm$1.7 \\
HD 128620 & 19.4$\pm$2.3 & 4.8$\pm$2.3 & 3.1$\pm$0.3 & 4.1$\pm$3.6 \\
HD 128621 & 16.7$\pm$0.7 & 8.6$\pm$8.1 & 3.1$\pm$0.1 & 6.0$\pm$4.5 \\
HD 136352 & 6.0$\pm$0.4 & 7.4$\pm$2.8 & 3.1$\pm$0.2 & 5.0$\pm$1.4 \\
HD 146233 & 6.2$\pm$0.1 & 6.9$\pm$2.6 & 3.2$\pm$0.1 & 4.2$\pm$1.2 \\
\enddata
\end{deluxetable}
%----------------------------------------------------------

\clearpage

\clearpage

\clearpage

\clearpage

\begin{figure} 
\plotone{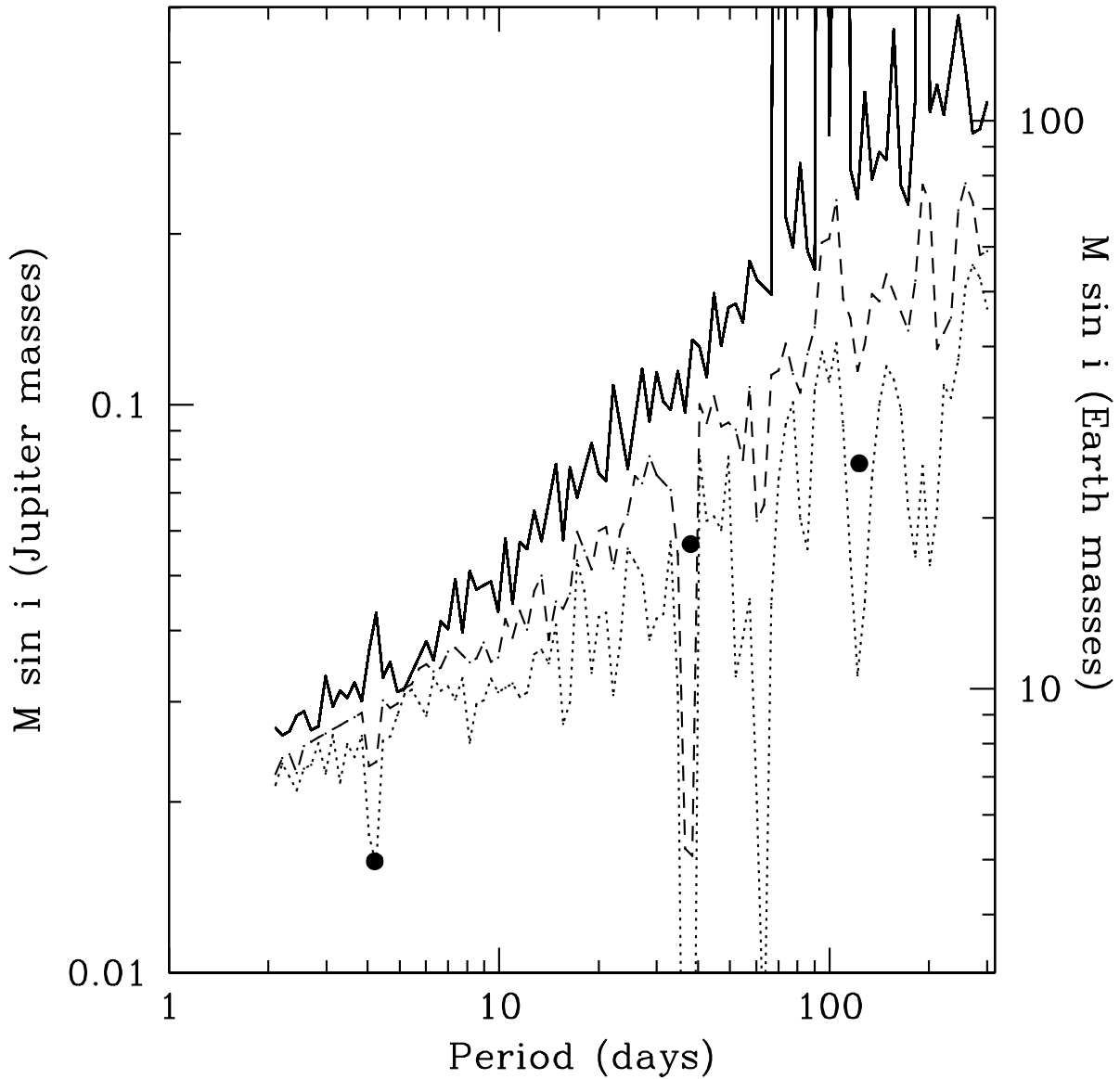} 
\caption{Detection limits for HD 115617 (61~Vir), which has recently 
been found to host three low-mass planets \citep{61vir}.  These planets 
were not accounted for in the detection-limit simulations presented 
here.  The lines represent recovery rates of 99\% (solid), 50\% 
(dashed), and 10\% (dotted).  The planets are shown here as filled 
circles.  The discovery of these planets was only possible with the 
inclusion of data from the Keck telescope and from a second 47-night 
Rocky Planet Search campaign in 2009 July/August.  In this work, we 
consider data from the AAT only, from 2005 June-2009 June. }
\label{61vir} 
\end{figure}

\end{document}